\documentclass[a4paper,fleqn,usenatbib]{mnras}

\usepackage{newtxtext,newtxmath}
\usepackage[T1]{fontenc}
\usepackage{ae,aecompl}

\usepackage[normalem]{ulem}
\usepackage{xfrac}
\usepackage{float}
\usepackage{epsfig}
\usepackage{footmisc}
\usepackage{multirow}
\usepackage{color}
\usepackage{graphicx}
\usepackage{longtable}
\usepackage{comment}
\usepackage{caption}
\usepackage{amsmath}
\usepackage{mathtools}
\usepackage{threeparttablex}
\usepackage[usenames,dvipsnames]{xcolor}

\def\um{$\mu$m}

\def\z{redshift}

\def\zphot{$z_{\mathrm{phot}}$}
\def\zspec{$z_{\mathrm{spec}}$}

\def\msol{$M_{\odot}$}

\def\lir{$L_{\rm IR}$}
\def\luv{$L_{\rm UV}$}

\def\rf{rest-frame}

\def\logoverdens{log($1 + \delta_{\rm gal}$)}

\title[SFR$-$density relations at $z \sim 0.9$]{Conditional Quenching: A detailed look at the SFR$-$Density Relation at $z\sim0.9$ from ORELSE}

\author[A. R. Tomczak et al.]
{\parbox{\textwidth}{Adam R. Tomczak$^{1}$\thanks{E-mail: atomczak724@gmail.com},
Brian C. Lemaux$^{1}$,
Lori M. Lubin$^{1}$,
Debora Pelliccia$^{1}$,
Lu Shen$^{1}$,
Roy R. Gal$^{2}$,
Denise Hung$^{2}$,
Dale D. Kocevski$^{3}$,
Olivier Le F\`evre$^{4}$,
Simona Mei$^{5,6,7}$,
Nicholas Rumbaugh$^{8}$,
Gordon K. Squires$^{9}$,
Po-Feng Wu$^{10}$\\
}\\\\
$^1$ Department of Physics, University of California, Davis, One Shields Ave., Davis, CA 95616, USA\\
$^2$ University of Hawai'i, Institute for Astronomy, 2680 Woodlawn Drive, HI 96822, USA\\
$^3$ Department of Physics and Astronomy, Colby College, Waterville, ME 04901, USA\\
$^4$ Aix Marseille Universit\'e, CNRS, LAM (Laboratoire d'Astrophysique de Marseille) UMR 7326, 13388, Marseille, France\\
$^5$ University of Paris Denis Diderot, University of Paris Sorbonne Cit\'e (PSC), 75205 Paris Cedex 13, France\\
$^6$ Sorbonne Universit\'{e}, Observatoire de Paris, Universit\'{e} PSL, CNRS, LERMA, F-75014, Paris, France\\
$^7$ Jet Propulsion Laboratory, Cahill Center for Astronomy \& Astrophysics, California Institute of Technology, 4800 Oak Grove Drive, Pasadena, California, USA\\
$^8$ National Center for Supercomputing Applications, University of Illinois, 1205 West Clark St., Urbana, IL 61801, USA\\
$^{9}$ Spitzer Science Center, California Institute of Technology, M/S 220-6, 1200 E. California Blvd., Pasadena, CA 91125, USA\\
$^{10}$ Max-Planck-Institut f\"{u}r Astronomie, K\"{o}nigstuhl 17, D-69117 Heidelberg, Germany\\
}

\date{Accepted XXX. Received YYY; in original form ZZZ}

\pubyear{2018}

\begin{document}
\label{firstpage}
\pagerange{\pageref{firstpage}--\pageref{lastpage}}
\maketitle

% Abstract of the paper
\begin{abstract}
We present a study of the star-formation rate (SFR)-density relation at $z \sim 0.9$ using data drawn from the Observations of Redshift Evolution in Large Scale Environments (ORELSE) survey.
%ORELSE targets a subset of the most massive ($\gtrsim 10^{14.4} M_{\odot}$) known large scale structures (LSSs) at $0.7 \leq z \leq 1.26$ with extensive spectroscopic coverage paving the way for detailed studies of galaxy evolution in the highest-density environments at these redshifts.
We find that SFR does depend on environment, but only for intermediate-stellar mass galaxies ($10^{10.1} < M_* / M_{\odot} < 10^{10.8}$) wherein the median SFR at the highest densities is $0.2-0.3$ dex less than at lower densities at a significance of $4\sigma$.
Interestingly, mass does not drive SFR; galaxies that are more/less massive have SFRs that vary at most by $\approx20\%$ across all environments showing no statistically significant dependence.
We further split galaxies into low-redshift ($z \sim 0.8$) and high-redshift ($z \sim 1.05$) subsamples and observe nearly identical behavior.
We devise a simple toy model to explore possible star-formation histories (SFHs) for galaxies evolving between these redshifts.
The key assumption in this model is that star-forming galaxies in a given environment-stellar mass bin can be described as a superposition of two exponential timescales (${\rm SFR} \propto e^{-t / \tau}$): a long$-\tau$ timescale with $\tau = 4$ Gyr to simulate ``normal'' star-forming galaxies, and a short$-\tau$ timescale with free $\tau$ (between $0.3 \leq \tau / {\rm Gyr} \leq 2$) to simulate galaxies on a quenching trajectory.
In general we find that galaxies residing in low/high environmental densities are more heavily weighted to the long$-\tau$/short$-\tau$ pathways respectively, which we argue is a signature of environmental quenching.
Furthermore, for intermediate-stellar mass galaxies this transition begins at intermediate-density environments suggesting that environmental quenching is relevant in group-like halos and/or cluster infall regions.
\end{abstract}

% Select between one and six entries from the list of approved keywords.
% Don't make up new ones.
\begin{keywords}
galaxies: evolution -- galaxies: groups: general -- galaxies: clusters: general -- techniques: photometric -- techniques: spectroscopic
\end{keywords}

%%%%%%%%%%%%%%%%%%%%%%%%%%%%%%%%%%%%%%%%%%%%%%%%%%
%%%%%%%%%%%%%%%%% BODY OF PAPER %%%%%%%%%%%%%%%%%%
%%%%%%%%%%%%%%%%%%%%%%%%%%%%%%%%%%%%%%%%%%%%%%%%%%

\section{Introduction}

One of the most sought after topics in the field galaxy evolution is determining {\it how galactic environment regulates star-formation in galaxies} and {\it how this changes over cosmic history}.
Efforts over the past several decades have found that there is no simple solution to these questions, but have instead revealed that there are many nuances that govern this aspect of galaxy evolution.
This has led many investigators to take a reductionist approach to the subject by studying general trends in galaxy properties as a function of environment one by one.
To cite a few examples from the local universe, galaxies in higher-density environments have been observed to more frequently have elliptical morphologies \citep[e.g.,][]{Dressler1980}, have older stellar populations \citep[e.g.,][]{Smith2006, Cooper2010a}, have redder colours and higher stellar masses \citep[e.g.,][]{Hogg2004, Kauffmann2004}, and lower levels of star-formation \citep[e.g.,][]{Lewis2002, Gomez2003}.
While these various properties can be examined independently they are still interdependent and correlated with each other.
In most cases, the genesis of new stellar populations via star-formation is at the root as the rate of star-formation in a galaxy has a direct influence on its stellar mass, mean stellar age, and colours.
Therefore, in this study we will focus on the latter of these, referred to as the star-formation rate (SFR)-density relation.

While this picture of ``red and dead'' galaxies in high-density environments has been established in the local universe many studies over roughly the past decade have pushed towards higher redshifts to investigate how this relation has evolved over cosmic history.
At $z \gtrsim 1.5$ several studies targeting galaxy overdensities have found evidence for cases where the SFRs of member galaxies are comparable or enhanced relative to field galaxies \citep[e.g.,][]{Tran2010, Strazzullo2013, Santos2015, Webb2015, Wang2016} indicating that environment did not always strictly act to quench star-formation.
These observations mainly pertain to cluster cores, in other works cluster members have shown systematically lower SFRs than similar mass counterparts among field galaxies at these redshifts \citep{Noirot2018}.
Unfortunately, however, understanding the nature of a transition in the SFR-density relation and the exact epoch at which it occurs is still debated with some studies finding positive correlations, some finding negative correlations, and others no significant correlation at intervening redshifts \citep[e.g.,][]{Elbaz2007, Cooper2008, Popesso2011, Patel2011, Grutzbauch2011, Muzzin2012, Koyama2013, Ziparo2014, Lin2014, Darvish2016}.
At least part of the tension among these works can be attributed to systematic differences within datasets and procedures such as SFR indicators (e.g. dust-corrected [OII], far-infrared), range and definitions of galaxy environment, and contamination from active galactic nuclei (AGN).
Furthermore, due to the difficulty in acquiring high quality data (particularly spectroscopy) for large enough samples to perform statistical tests many of these studies do not have the luxury of creating subsamples but instead must make measurements and statements on the entire galaxy sample they have data for.

In this work we make use of the Observations of Redshift Evolution in Large Scale Environments Survey \citep[ORELSE:][]{Lubin2009} to explore the SFR-density relation at $z \sim 0.9$.
ORELSE is an extensive photometric and spectroscopic survey of 16 of the most massive large scale structures (LSSs) known at $0.7 \leq z \leq 1.26$.
With hundreds of high-quality spectroscopic redshifts per field these data make it possible to accurately map out the three-dimensional density field of each LSS, which reveal a large diversity of substructure \citep[see][]{Lemaux2017, Shen2017, Rumbaugh2017}.
Combining with these data far-infrared imaging from {\it Spitzer}/Multiband Imaging Photometer for {\it Spitzer} \citep[MIPS:][]{Rieke2004} for measuring SFRs, allows us to carefully explore the variation of the SFR-density relation across the full range of environmental densities at these redshifts.

This paper is laid out as follows.
In Section \ref{sec:dataandmethods} we discuss the data and sample selection, spectral energy distribution (SED)-fitting, and how environment and SFRs are estimated.
Our main results regarding the SFR-density relation are presented in Section \ref{sec:analysis} where we look at the SFR-$M_*$ relation in bins of environmental density as well as specific star-formation rate ($\rm SSFR \equiv SFR / {\it M}_*$) as a function of environment and stellar mass.
To explain these observations using a physical picture we devise a toy model which we describe and discuss the results and interpretation of in section \ref{sec:simulation}.
We conclude with a summary in Section \ref{sec:conclusions}.
All magnitudes are presented in the Absolute Bolometric system \citep[AB:][]{Oke1983}.
Throughout this paper we adopt a standard $\Lambda$CDM cosmology with $\Omega_{\mathrm{M}} = 0.3$, $\Omega_{\mathrm{\Lambda}} = 0.7$, and $H_0 = 70$ km s$^{-1}$ Mpc$^{-1}$.

% ~~~~~~~~~~~~~~~~~~~~~~~~~~~~~~~~~~~~~~~~~~~~~~~~~~~~~~~~~~~~~
% ~~~~~~~~~~~~~~~~~~~~~~~~~~~~~ TABLE ~~~~~~~~~~~~~~~~~~~~~~~~~~
% ~~~~~~~~~~~~~~~~~~~~~~~~~~~~~~~~~~~~~~~~~~~~~~~~~~~~~~~~~~~~~

\begin{table}
    \begin{center}
    \caption{Large Scale Structures}
    \label{tab:lss}

    \begin{tabular}{lcccc}

    \hline \\[-5mm]  
    \hline \\[-3mm]  

    Name & R.A.$^a$ & Dec.$^a$ & $\langle$\zspec$\rangle$$^b$ & log($M_{\rm vir}$)$^c$ \\
     & [J2000] & [J2000] & & [M$_{\odot}$] \\[-0.5mm]

    \hline \\[-2mm]

    RXJ1221  &   185.3521   &   94.3036   &   0.700   &   14.7  \\
    RCS0224  &   36.14120   &   -0.0394   &   0.772   &   15.0  \\
    CL1350   &   207.7021   &   60.1186   &   0.804   &   14.7  \\
    RXJ1716  &   259.2016   &   67.1392   &   0.813   &   15.3  \\
    RXJ1821  &   275.3845   &   68.4658   &   0.818   &   15.2  \\
    SG0023   &     5.9675   &    4.3853   &   0.845   &   14.4  \\
    SC1604   &   241.1409   &   43.3539   &   0.898   &   15.4  \\
    CL1429   &   217.2767   &   42.6861   &   0.987   &   14.8  \\
    XLSS005  &    36.7904   &   -4.3014   &   1.056   &   14.5  \\
    SC0910   &   137.5983   &   54.3419   &   1.110   &   15.0  \\
    RXJ1053  &   163.4158   &   57.5883   &   1.140   &   14.8  \\
    SC0849   &   132.2333   &   44.8711   &   1.261   &   15.0  \\[0.5mm]

    \hline

    \end{tabular}
    \end{center}

    $^a$ Central coordinates of the main structure in each field. \\
    $^b$ Mean spectroscopic redshift of the main structure in each field. \\
    $^c$ Total virial mass of the main structure in each field.
    In many systems we identify multiple galaxy groups and/or clusters.
    For these cases we show the sum of the virial masses of all substructures.
\end{table}

\begin{figure*}
    \includegraphics[width=2\columnwidth]{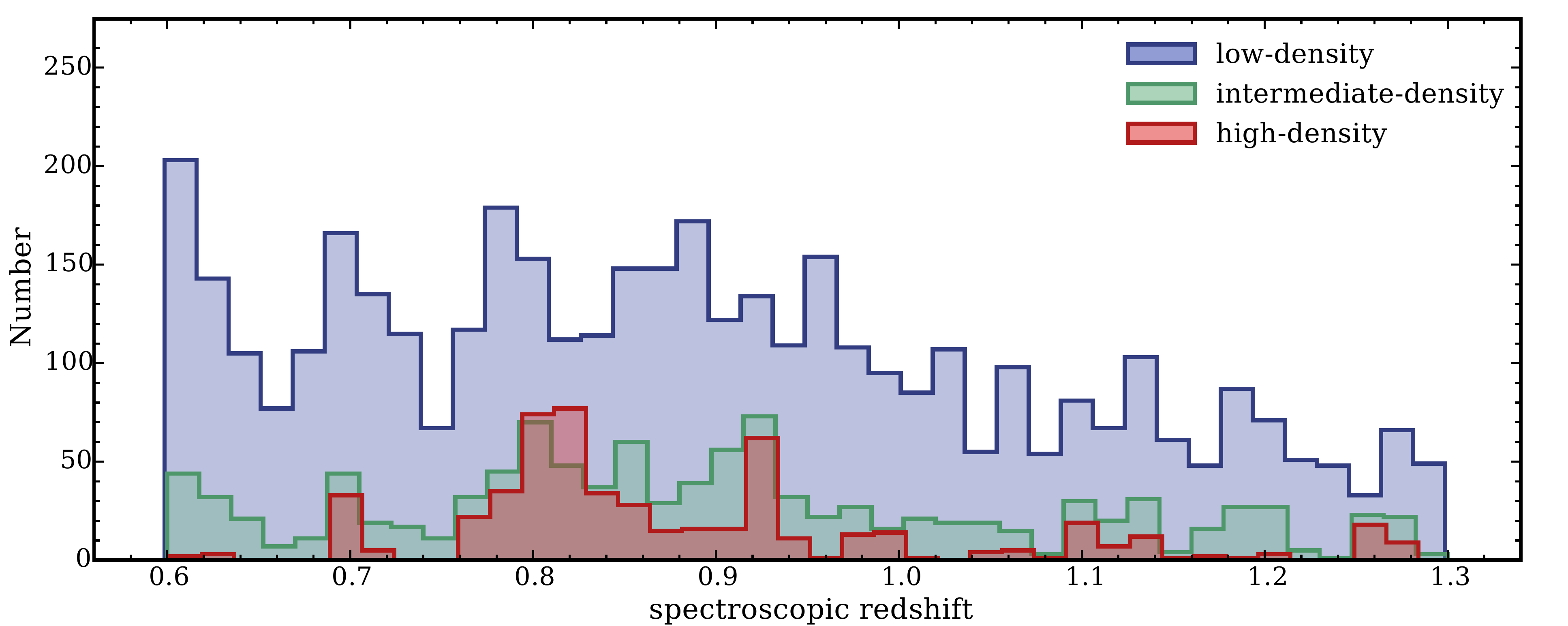}
    \caption{
    Spectroscopic redshift (\zspec ) distributions of all galaxies in the three environmental density bins considered in this work: low-density (blue), intermediate-density (green), and high-density (red).
    The total numbers of all galaxies in these samples are 4147, 1078, and 545 respectively.
    After applying the star-forming selection criteria, as described in Section \ref{sec:uvj}, the total numbers of star-forming galaxies in these samples are 3451, 729, and 229 respectively.
    The median \zspec\ of these samples are 0.88, 0.89, and 0.84 respectively.
    While these redshift distributions are slightly disparate they do not correspond to significant differences in the observed redshift-evolution of the mean SFR at fixed stellar mass \citep[e.g.,][]{Speagle2014}.
    Nonetheless, we correct for these slight differences in redshift as described in Section \ref{sec:analysis}.
    }
    \label{fig:zhistograms}
\end{figure*}

\section{Data and Methods}
\label{sec:dataandmethods}

Data used in this work are drawn from the Observations of Redshift Evolution in Large Scale Environments survey \citep[ORELSE:][]{Lubin2009}.
ORELSE is a panchromatic photometric and spectroscopic campaign of massive large scale structures (LSSs) in the redshift range $0.6 < z < 1.3$.
The spectroscopic portion has yielded thousands of high-quality spectra across 16 well-known galaxy cluster fields (400$-$1000 per field).
In this study we only utilize fields for which we have far-infrared imaging from the {\it Spitzer} Space Telescope necessary for the estimation of SFRs, reducing the number to 12.
Table \ref{tab:lss} lists basic properties for this subset of large scale structure fields.
For a more intensive discussion of photometric and spectroscopic data we refer the reader to \citet{Tomczak2017} as here we provide only a brief summary.

\subsection{Photometry}

For the photometric component, deep multi-wavelength imaging are drawn archival as well as targeted observations from a wide range of facilities including Subaru, Palomar, CFHT, UKIRT, and {\it Spitzer}.
As a result, the photometry is heterogeneous as the exact combination of bandpasses varies from field to field.
Nevertheless, at a minimum each field has imaging in at least nine bandpasses spanning 0.4$-$4.5 \um , which most commonly include $B, V, R, I, Z, J, K, [3.6],$ and [4.5].
Optical/Near-Infrared photometry ($B$ through $K$ filters) is measured in fixed circular apertures on images smoothed to the seeing of the largest point spread function (PSF) within each field.
However, for {\it Spitzer}/InfraRed Array Camera \citep[IRAC:][]{Fazio2004} imaging we use {\tt T-PHOT} \citep{Merlin2015}, a software package designed to extract accurate photometry in images where crowding is an issue due to large PSFs.
Table \ref{tab:photometry1} in Appendix \ref{appendix} lists all optical/near-IR photometry, including facilities and instruments as well as depth estimates for all fields considered in this study.

In addition to these data, far-infrared imaging in the 24 \um\ bandpass from {\it Spitzer}/Multiband Imaging Photometer for {\it Spitzer} \citep[MIPS:][]{Rieke2004} were obtained.
The procedure for processing and extracting photometry in these data are similar to what was done for the {\it Spitzer}/IRAC data as described in Section 2.1 of \citet{Tomczak2017}.
Data were downloaded from the {\it Spitzer} Heritage Archive\footnote{\url{http://sha.ipac.caltech.edu/applications/Spitzer/SHA/}} and reduced using the {\tt MOPEX} software package \citep{Makovoz2006}.
For each field separately, MOPEX is run on the individual corrected basic calibrated data (cBCD) frames.
First, overlapping frames are background-matched using the {\tt overlap.pl} routine which calculates additive offsets among exposures.
Next, mosaicing is performed using the {\tt mosaic.pl} routine which performs interpolation, outlier rejection, and coaddition of frames.
This step also produces coverage and uncertainty maps that show the number of frames and uncertainty per pixel respectively, the latter of which is used with {\tt T-PHOT} for extracting photometry.

To estimate the depth of these 24\um\ data we employ the following procedure.
In each image we measure the fluxes in 1000 randomly placed apertures of diameter $d = 10.4$\arcsec\ (roughly 2$\times$ the PSF full-width-half-max) in empty sky locations.
Objects are masked using a segmentation map generated using the Source Extractor software \citep[SExtractor:][]{Bertin1996}.
We then fit a Gaussian to the distribution of these fluxes and adopt the standard deviation as an estimate of the 1$\sigma$ source detection limit.
Finally, we apply an aperture correction to account for flux that falls outside of the aperture defined above, which we find corresponds to roughly a 1.7$\times$ correction factor.
Across all fields listed in Table \ref{tab:lss} we estimate 24\um imaging depths between 30$-$115 $\mu$Jy, which roughly correspond to 4$-$13 $M_{\odot}$/yr in terms of SFR$_{\rm IR}$ (see Section \ref{sec:sfrs}).
Note that this does not limit the present study to these high SFR limits because of methods described in later sections.

\subsection{Spectroscopy}

The majority of spectroscopic data used in this study are taken from the ORELSE survey which utilized \ion{Keck}{2}/DEep Imaging Multi-Object Spectrograph (DEIMOS; \citealt{Faber2003}).
For a full overview of the spectroscopic targeting and observations for ORELSE we refer the reader to \citet{Gal2008}, \citet{Lubin2009}, \citet{Lemaux2009}, and \citet{Rumbaugh2017}

To summarize, targets for these observations were preselected based on having observed-frame ($r^{\prime}-i^{\prime}$) and ($i^{\prime}-z^{\prime}$) colours consistent with the known LSS redshifts.
Priority was given to objects on or near the red sequence with bluer objects receiving progressively lower priorities.
Objects fainter than $i^{\prime}=24.5$ were generally not targeted, although this was not a strict limit. 
Occasionally multiple objects would fall into a single slit leading and yield a serendipitous redshift, which account for roughly 3.6\%\ of all spectroscopic redshifts.
It is worth mentioning that although redder objects were preferentially targeted the majority of spectroscopic targets have blue colours due to the strictness of the exact colour cuts and lower numbers of red objects.
Indeed, the final spectroscopic sample has been shown to be broadly representative of the underlying galaxy population in terms of
stellar mass ($>$10$^{10.2}$ $M_{\odot}$) and colours ($M_{NUV} - M_{r} > 2.5$) \citep[see Appendix B of][]{Shen2017}.

In addition to the DEIMOS observations, a notable number of spectroscopic targets (2273, roughly 39\%) used in this work are drawn from other literature studies \citep{Oke1998, Gal2004, Tanaka2008, Mei2012} which utilized different telescopes and instruments. 
Of these additional targets, the majority (2043) come from the VIMOS VLT Deep Survey \citep[VVDS:][]{LeFevre2013} located entirely in the XLSS005 field.
For galaxies targeted by more than one survey we adopt the redshift with the higher confidence, with ties going to ORELSE.
Additional information regarding spectroscopic observations in this field can be found in \citet{Lemaux2014}.

\subsection{Spectral Energy Distribution Fitting}
\label{sec:SEDfitting}

We perform Spectral Energy Distribution (SED) fitting in a two-stage process.
First, we utilize the code Easy and Accurate Redshifts from Yale \citep[{\tt EAZY}:][]{Brammer2008} to estimate photometric redshifts for galaxies that lack spectroscopic redshifts.
{\tt EAZY} is able to simultaneously estimate rest-frame photometry by convolving SED fits with requested filter transmission curves using either the best-fit \zphot\ or a provided \zspec\ (when available).
We use this feature to obtain rest-frame $(U - V)$ and $(V - J)$ broadband colours for classifying star-formation activity in galaxies, as well as rest-frame 2800\AA\ luminosities (${L}_{\nu, 2800}$) for estimating unobscured star-formation rates.

In the second stage of SED fitting we employ the code Fitting and Assessment of Synthetic Templates \citep[{\tt FAST}:][]{Kriek2009} to estimate stellar masses as well as other properties of the stellar populations of galaxies.
For this we adopt the stellar population synthesis models presented by \citet{Bruzual2003} with solar metallicity, following a delayed exponential star-formation history of the form ${\rm SFR} \propto t \times e^{-t / \tau}$, assuming a \citet{Chabrier2003} stellar initial mass function, and allowing for dust attenuation following the \citet{Calzetti2000} extinction law.
Similar to the FourStar Galaxy evolution survey \citep[ZFOURGE:][]{Straatman2016} we allow log($\tau / {\rm yr}$) to range between 7 and 10 in steps of 0.2, log(${\rm age / yr}$) to range between 7.5 and 10.1 in steps of 0.1, and $A_V$ to range between 0 and 4 in steps of 0.1.
For a more thorough explanation of the SED fitting protocol used for this study see Section 2.3 of \citet{Tomczak2017}.

\subsection{Sample Selection}

Galaxies were selected from 12 unique fields, each hosting a known massive large scale structure in the redshift range [0.7$-$1.26] (See Table \ref{tab:lss}).
Across all fields we select galaxies with secure spectroscopic redshifts between $0.6 < z_{\rm spec} < 1.3$, ensuring that we sample galaxies over the full extent of each LSS as well as including those in intermediate-density (cluster outskirts, filaments, groups, etc.) and low-density field environments.
Although the estimated stellar mass completeness limits for these fields range between $10^{9} - 10^{10.5}$ \msol\ at these redshifts \citep[see][]{Tomczak2017} we do not apply a stellar mass cut to this initial sample.
Even though we will be incomplete in terms of numbers, it does not necessarily or immediately follow that our galaxy sample will be biased in terms of SFR at low stellar masses.

Additionally, we reject galaxies that have been flagged as having poor photometry based on the methods described in Section 2.3 of \citet{Tomczak2017}.
This ``use'' flag is designed to identify objects that are poorly detected, saturated, have catastrophic SED fits, and/or likely to be foreground stars.
In general, these criteria predominantly apply to photometrically-selected objects because spectroscopic targeting, for example, is designed to avoid saturated objects and foreground stars.
Nevertheless, roughly 1.5\%\ of galaxies with high-quality spectroscopic redshifts are rejected based on catastrophic SED fits (defined as having $\chi^2_{\rm reduced} > 10$).
We also reject galaxies that have been identified as possibly having their optical/NIR SEDs dominated by an active galactic nucleus (AGN) based on X-ray emission \citep{Rumbaugh2017} or radio emission \citep{Shen2017}.
It is important to note, however, that only three fields used in this study (RXJ1821, SG0023, SC1604) currently have sufficient radio observations for this classification.

Finally, we require that galaxies have coverage in {\it Spitzer}/MIPS 24\um\ imaging for estimation of star-formation rates (see Section \ref{sec:sfrs}).
In Figure \ref{fig:zhistograms} we show the spectroscopic redshift distribution of our full sample of 5770 galaxies split into three environmental density bins defined in Section \ref{sec:analysis}.

\begin{figure*}
    \includegraphics[width=2\columnwidth]{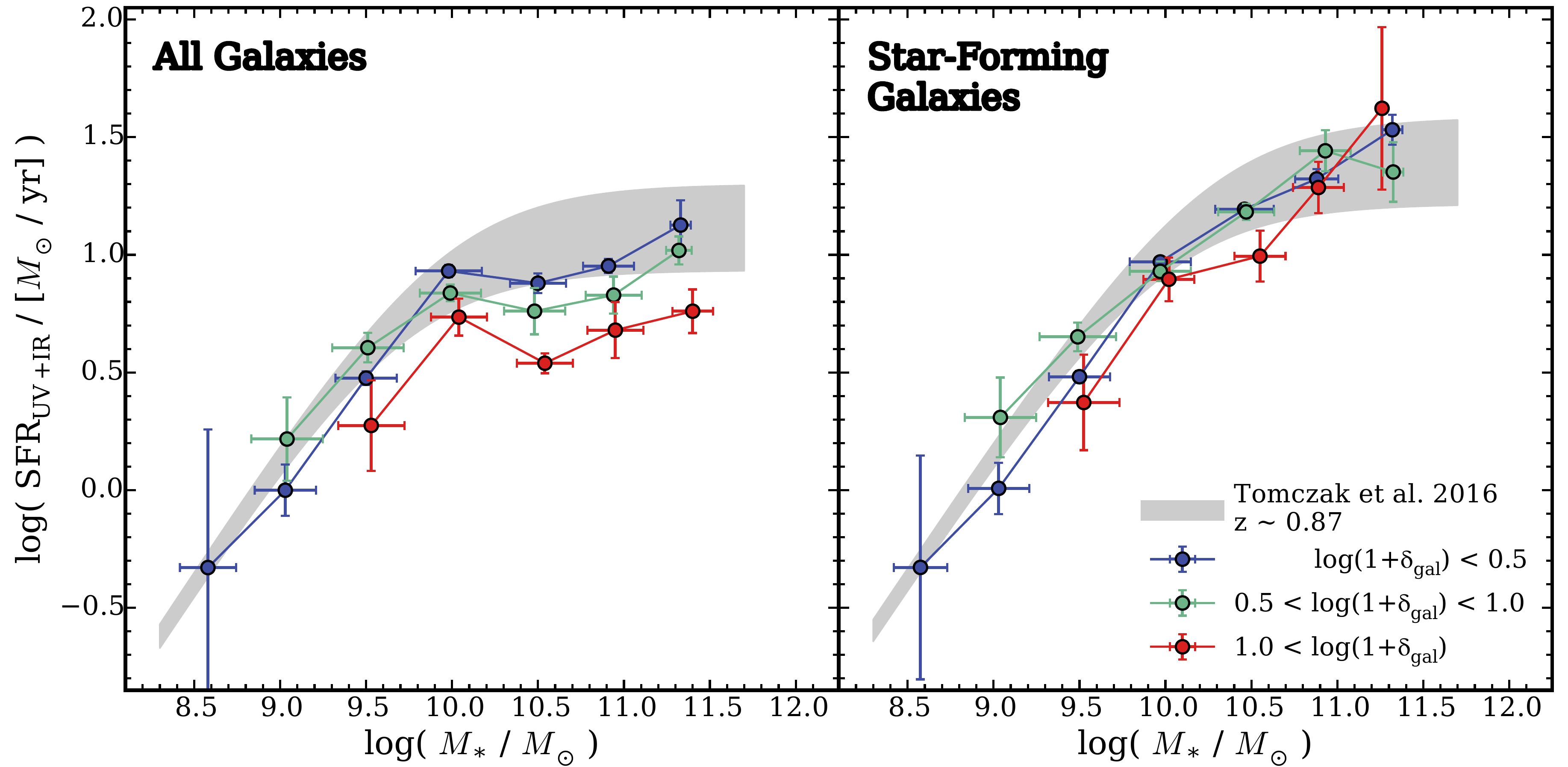}
    \caption{
    Median star-formation rate (SFR) as a function of stellar mass split into three bins of environmental density for all ({\it left}) and star-forming ({\it right}) galaxies.
    Errorbars indicate the 1$\sigma$ scatter on the median SFR of 1000 bootstrapped realizations of each subsample.
    For reference, the gray shaded regions show the field SFR-$M_*$ relations from ZFOURGE \citep{Tomczak2016} around the median redshift the full sample studied here, which should be (and indeed are) comparable to our low-density sample (blue points).
    For all galaxies the median SFR at fixed $M_*$ systematically decreases from low- to high-density environments at all log($M_*$/$M_{\odot}$) $\gtrsim 9.5$.
    However, when considering only star-forming galaxies (i.e. excluding quenched galaxies) this trend is suppressed, although still tenable at $\sim$10$^{10.5}$ $M_{\odot}$.
    This suggests that an increased prevalence of quiescent galaxies in higher-density environments is largely, albeit not entirely, responsible for the observed anticorrelation between SFR and density at fixed $M_*$.
    }
    \label{fig:sfrmass_relations}
\end{figure*}

\subsection{Estimating Local Environment}
\label{sec:environment}

We estimate local environmental densities of galaxies using a Voronoi Monte Carlo algorithm.
For a full discussion of the procedure and tests of its performance we refer the reader to sections 3.2 and 3.3 of \citet{Lemaux2017} and \citet{Tomczak2017} respectively.
Additionally, in a forthcoming paper we will present results from a series of rigorous tests aimed at describing the efficacy of tracing large scale structure using this technique (Hung et al. in preparation).

Briefly, galaxies are partitioned into thin \z\ slices, each spanning 3000 km/s in velocity space and overlapping by 1500 km/s from slice to slice.
For each \z\ slice a projected 2-dimensional Voronoi tessellation is calculated using the positions of galaxies that fall within it.
Local environmental {\it density} is defined as the inverse of the area of a Voronoi cell multiplied by the square of the angular diameter distance at the corresponding mean \z\ of the slice.
These maps are then projected onto a two dimensional grid of pixels of size $75 \times 75$ proper kpc.
The local environmental {\it overdensity} at pixel $(i, j)$ is defined as:

\begin{equation}
\hspace{18mm}
\mathrm{log}(1 + \delta_{\mathrm{gal}}) \equiv \mathrm{log} \bigg( \; 1 + \frac{\Sigma_{i, j} - \tilde{\Sigma}}{\tilde{\Sigma}} \; \bigg)
\end{equation}

\noindent
where $\Sigma_{i, j}$ is the density at pixel $(i, j)$ and $\tilde{\Sigma}$ is the median density of all pixels where the map is reliable (i.e. coverage in nearly all images and not near the edge of the detection image).
The local (over)density for a given galaxy is set by the closest pixel in proximity in the transverse and redshift dimensions.
{\color{black}
We find this metric to be strongly concordant with more traditional metrics such as, e.g., the projected radial distances of galaxies relative to the centroid of their nearest group or cluster.

A detailed comparative study performed by \citet{Darvish2015} tests a variety of environmental density metrics against simulated galaxy catalogs.
The authors find that the Voronoi tessellation performs among the best at recovering the true environmental densities.
We refer the reader to this and \citet{Muldrew2012} for thorough tests and discussions of methods for estimating galaxy environment, as well as Hung et al. in prep for tests related to our methodology of estimating galaxy environment.
}

\subsection{Classifying Star-Forming and Quiescent Galaxies}
\label{sec:uvj}

We facilitate this work by separating galaxies into two classes, star-forming and quiescent, for the purposes of studying the SFRs of actively star-forming galaxies as well as quantifying the fraction of galaxies that are quenched.
For this task we make use of the rest-frame $(U-V)$ and $(V-J)$ colours estimated by {\tt EAZY} (Section \ref{sec:SEDfitting}).
This pair of broadband colours has been shown to be effective at disambiguating reddening due to aging versus dust attenuation, allowing for a robust method of classifying star-forming and quenched galaxies \citep[e.g.,][]{Labbe2005, Wuyts2007, Williams2009}.
Galaxies are classified as quiescent, and subsequently removed from our sample, if they satisfy these criteria: $(U - V) > 1.3 \; \cap \; (U - V) > 0.88 (V - J) + 0.59$.

\begin{table*}
    \begin{center}
    \caption{Median SFR-$M_*$ Relations}
    \label{tab:sfrmass}

    \begin{tabular}{c r rrc rrc}

    \multicolumn{8}{c}{Full Sample ($0.6 < z < 1.3$)} \\[-1mm]
    \hline \\[-5mm]
    \hline \\[-3mm]

    Overdensity  &   \multicolumn{1}{c}{log($M_*$)}  &  \multicolumn{1}{c}{log$(\widetilde{M_*})_{\rm ALL}$$^a$}  &  \multicolumn{1}{c}{log$(\widetilde{\rm SFR})_{\rm ALL}$$^b$}  &  $N_{\rm ALL}$$^c$  &  \multicolumn{1}{c}{log$(\widetilde{M_*})_{\rm SF}$$^a$}  &  \multicolumn{1}{c}{log$(\widetilde{\rm SFR})_{\rm SF}$$^b$}  &  $N_{\rm SF}$$^c$ \\
        Bin      & \multicolumn{1}{c}{[$M_{\odot}$]} & \multicolumn{1}{c}{[$M_{\odot}$]} & \multicolumn{1}{c}{[$M_{\odot}$/yr]} & & \multicolumn{1}{c}{[$M_{\odot}$]} & \multicolumn{1}{c}{[$M_{\odot}$/yr]} & \\
    \hline \\[-3mm]
      low       & $ 8.25 -  8.75$  &  $ 8.58 \pm 0.16$  &  $-0.33 \pm 0.59$  &  241 &
                                       $ 8.57 \pm 0.16$  &  $-0.33 \pm 0.48$  &  240 \\
    density     & $ 8.75 -  9.25$  &  $ 9.03 \pm 0.18$  &  $-0.00 \pm 0.11$  &  779 &
                                       $ 9.03 \pm 0.18$  &  $0.01 \pm 0.11$  &  773 \\
                 & $ 9.25 -  9.75$  &  $ 9.50 \pm 0.18$  &  $0.48 \pm 0.03$  &  1016 &
                                       $ 9.50 \pm 0.18$  &  $0.48 \pm 0.03$  &  993 \\
                 & $ 9.75 - 10.25$  &  $ 9.98 \pm 0.19$  &  $0.93 \pm 0.03$  &  866 &
                                       $ 9.97 \pm 0.18$  &  $0.97 \pm 0.02$  &  780 \\
                 & $10.25 - 10.75$  &  $10.50 \pm 0.16$  &  $0.88 \pm 0.04$  &  759 &
                                       $10.46 \pm 0.17$  &  $1.19 \pm 0.02$  &  447 \\
                 & $10.75 - 11.25$  &  $10.91 \pm 0.15$  &  $0.95 \pm 0.03$  &  418 &
                                       $10.88 \pm 0.13$  &  $1.32 \pm 0.04$  &  184 \\
                 & $11.25 - 11.75$  &  $11.33 \pm 0.06$  &  $1.13 \pm 0.10$  &   51 &
                                       $11.32 \pm 0.06$  &  $1.53 \pm 0.06$  &   17 \\
    \hline \\[-3mm]
  intermediate      & $ 8.75 -  9.25$  &  $ 9.04 \pm 0.21$  &  $0.22 \pm 0.18$  &  127 &
                                       $ 9.04 \pm 0.21$  &  $0.31 \pm 0.17$  &  125 \\
      density    & $ 9.25 -  9.75$  &  $ 9.51 \pm 0.21$  &  $0.61 \pm 0.06$  &  179 &
                                       $ 9.49 \pm 0.22$  &  $0.65 \pm 0.06$  &  168 \\
                 & $ 9.75 - 10.25$  &  $ 9.99 \pm 0.18$  &  $0.84 \pm 0.04$  &  213 &
                                       $ 9.97 \pm 0.18$  &  $0.93 \pm 0.04$  &  166 \\
                 & $10.25 - 10.75$  &  $10.48 \pm 0.18$  &  $0.76 \pm 0.10$  &  303 &
                                       $10.47 \pm 0.16$  &  $1.18 \pm 0.03$  &  162 \\
                 & $10.75 - 11.25$  &  $10.94 \pm 0.16$  &  $0.83 \pm 0.08$  &  190 &
                                       $10.93 \pm 0.15$  &  $1.44 \pm 0.09$  &   67 \\
                 & $11.25 - 11.75$  &  $11.32 \pm 0.07$  &  $1.02 \pm 0.06$  &   33 &
                                       $11.32 \pm 0.06$  &  $1.35 \pm 0.13$  &    8 \\
    \hline \\[-3mm]
    high         & $ 9.25 -  9.75$  &  $ 9.53 \pm 0.19$  &  $0.27 \pm 0.19$  &   65 &
                                       $ 9.52 \pm 0.21$  &  $0.37 \pm 0.20$  &   50 \\
    density       & $ 9.75 - 10.25$  &  $10.04 \pm 0.16$  &  $0.74 \pm 0.08$  &  107 &
                                       $10.02 \pm 0.15$  &  $0.90 \pm 0.09$  &   63 \\
                 & $10.25 - 10.75$  &  $10.54 \pm 0.16$  &  $0.54 \pm 0.04$  &  169 &
                                       $10.55 \pm 0.15$  &  $0.99 \pm 0.11$  &   55 \\
                 & $10.75 - 11.25$  &  $10.95 \pm 0.16$  &  $0.68 \pm 0.12$  &  135 &
                                       $10.89 \pm 0.15$  &  $1.29 \pm 0.11$  &   30 \\
                 & $11.25 - 11.75$  &  $11.40 \pm 0.12$  &  $0.76 \pm 0.09$  &   38 &
                                       $11.26 \pm 0.01$  &  $1.62 \pm 0.34$  &    3 \\

    \hline \\[-5mm]

    \end{tabular}

    {\flushleft
    \hspace{15mm}
    $^a$ Median stellar mass of all and star-forming galaxies per bin.
    %Errors correspond to 1$\sigma$ NMAD scatter. \\
    Errors correspond to the normalized median absolute deviation (NMAD). \\
    \hspace{15mm}
    $^b$ Median log(SFR$_{\rm UV+IR}$) of all and star-forming galaxies per bin.
    Errors correspond to 1$\sigma$ scatter from 1000 bootstrap realizations. \\
    \hspace{15mm}
    $^c$ Number of all and star-forming galaxies per bin. \\
    }

    \end{center}

\end{table*}

\subsection{Estimating Star-Formation Rates}
\label{sec:sfrs}

Star-formation rates are estimated by adding contributions from obscured and unobscured young stellar populations as traced by IR and UV emission respectively.
We use the calibration presented by \citet{Bell2005} scaled to a \citet{Chabrier2003} IMF in this work:

\begin{equation}
{\rm SFR_{UV+IR}} \: [M_{\odot} / \mathrm{yr}]   =   1.09 \times 10^{-10} \: ( 2.2 \: {L}_{\rm UV}  +  {L}_{\rm IR} ) \: [{L_{\odot}}]
\end{equation} \\[-3mm]

\noindent
where \lir\ is the bolometric infrared luminosity (8$-$1000\um) and \luv $\, =$$\, 1.5 \, \nu \, {L}_{\nu, 2800}$ represents the \rf\ 1216$-$3000\AA\ luminosity, estimated using the methodology described in section \ref{sec:SEDfitting}.

Infrared luminosities are estimated by fitting the IR spectral template introduced by \citet{Wuyts2008} (hereafter W08) to our measured MIPS 24\um\ photometry.
This single template has been shown to accurately recover the average SFR of star-forming galaxies \citep{Muzzin2010, Tomczak2016}.
These fits are performed by first shifting the W08 template to the \zspec\ of a galaxy and then scaling to match the measured 24\um\ flux.
We then integrate between \rf\ 8$-$1000\um\ and treat this as the bolometric IR luminosity (\lir ).
The error of \lir\ is derived by propagating the error in the 24\um\ flux.

\section{Analysis}
\label{sec:analysis}

To compare galaxies as a function of environment we define three bins based on the local overdensity metric described in Section \ref{sec:environment}.
We refer to these subsamples as low-, intermediate-, and high-density which respectively contain galaxies having \logoverdens\ $< 0.5$, $0.5 <$ \logoverdens\ $< 1.0$, and \logoverdens\ $> 1.0$.
The number of total (star-forming) galaxies in each of these subsamples is 4147 (3451), 1078 (729), and 545 (229) respectively.
While the specific ranges of these bins are chosen somewhat arbitrary they do a reasonable job of separating galaxies based on the cumulative distribution of \logoverdens\ within the full ORELSE sample \citep[see Fig. 8 of][]{Shen2017}.
Spectroscopic redshift distributions are shown in Figure \ref{fig:zhistograms}, from which mild differences can bee seen.
Performing a two-sample Kolmogorov-Smirnov (KS) test for all pairings we find a maximum $p-$value of 0.00008, indicating that all three subsamples are statistically unique ($p < 0.05$).

Due to the fact that the mean SFR at fixed stellar mass evolves with redshift \citep[e.g.,][]{Speagle2014} there is a potential for bias when comparing the SFRs of galaxies among samples with different redshift distributions.
To account for this we make use of work from \citet{Tomczak2016} (hereafter T16) to derive corrections based on their parameterization of the mean SFR as a function of redshift and stellar mass: $\Psi_{\rm T16}(z, M_*)$.
We set the median spectroscopic redshift of the full sample as the ``fiducial'' redshift ($z_{\rm fiducial} = 0.87$) and calculate the median \zspec\ for each 2D bin of environment and stellar mass.
The correction factor is defined as the ratio between the SFR at the fiducial redshift, $\Psi_{\rm T16}(z_{\rm fiducial}, M_{*, i})$, and the SFR at the median \zspec\ of each environment-stellar mass bin, $\Psi_{\rm T16}(z_{\rm median}, M_{*, i})$, where $M_{*, i}$ is the central stellar mass of the bin.
Correction factors can be as large as 24\%\ but are typically low (4\% on average).

\begin{figure*}
    \includegraphics[width=2\columnwidth]{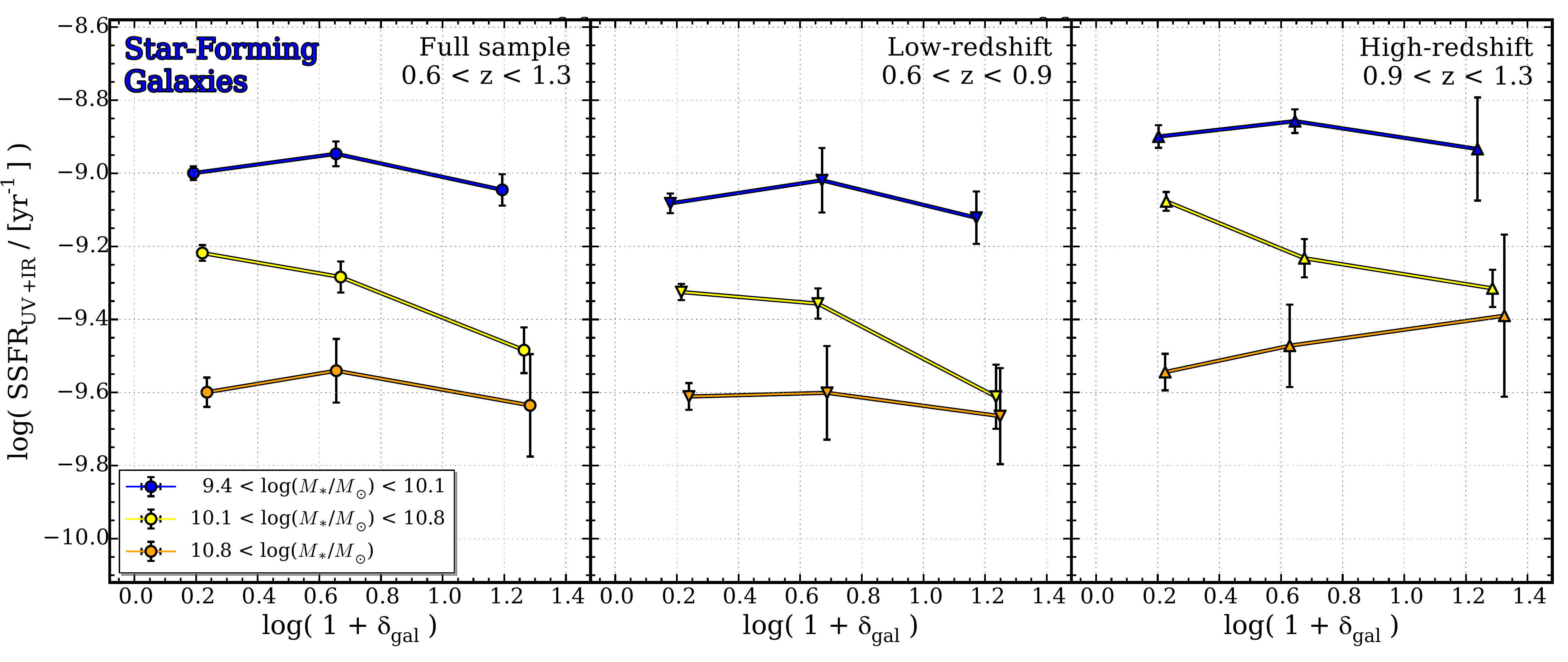}
    \caption{
    Median specific star-formation rate (SSFR) as a function of local environment for star-forming galaxies split into three stellar mass bins for the full sample ($0.6 < z < 1.3$), low-redshift sample ($0.6 < z < 0.9$), and high-redshift sample ($0.9 < z < 1.3$).
    Errorbars indicate the 1$\sigma$ scatter on the median SSFR of 1000 bootstrapped realizations of each subsample.
    For the low- and high-mass bins shown here SSFR is largely independent of environment.
    However, at intermediate stellar masses we observe a 0.3 dex decrease between the low- and high-density subsamples, significant at the $\geq 3 \sigma$ level in all cases, indicating that environment has a negative impact on the SFRs of these galaxies.
    Data from this figure can be found in Table \ref{tab:ssfrs}, as well as measurements split into low- and high-redshift subsamples ($0.6 < z < 0.9$ and $0.9 < z < 1.3$, respectively).
    }
    \label{fig:ssfrdensity_relations}
\end{figure*}

In Figure \ref{fig:sfrmass_relations} we plot the median, corrected SFR in bins of stellar mass for the three environment subsamples.
To estimate uncertainties we generate 1000 bootstrap realizations of each environment-stellar mass subsample and calculate the 1$\sigma$ scatter on the median SFR.
For reference we plot the range of the mean SFR-$M_*$ relation from T16 at $0.73 < z < 1.13$ (the central 1$\sigma$ spread of our sample).
In general we find a remarkably high level of concordance between the ZFOURGE field sample and our measurements for the low-density sample, which should be the most analogous to the ZFOURGE data in terms of environment.
When considering all galaxies we see a strong decrement in SFR with increasing environmental density consistently for galaxies with stellar masses $\gtrsim 10^{9.5}$ \msol.
However, this effect can be the result of galaxies having lower SFRs at fixed stellar mass in denser environments or it could be a reflection of an elevated fraction of quenched galaxies in denser environments.
To first order we can look into this by remeasuring these relations for galaxies that are classified as star-forming based on rest-frame broadband colours (see Section \ref{sec:uvj}).
If lower SFRs at fixed stellar mass in dense environments are primarily driven by increased numbers of quenched galaxies, then the SFR gradient should decrease or disappear when excluding quenched galaxies.
Indeed we do observe that the SFR-$M_*$ relation for star-forming galaxies (right panel of Figure \ref{fig:sfrmass_relations}) exhibits a much weaker dependence on environment, although, still appears relevant at $\sim 10^{10.5}$ $M_{\odot}$.
Data points as well as numbers of galaxies in each environment-mass bin are given in Table \ref{tab:sfrmass}.

To look at this more closely, we plot in the left panel of Figure \ref{fig:ssfrdensity_relations} the median, corrected specific star-formation rate (${\rm SSFR \equiv SFR} / M_*$) as a function of local overdensity for star-forming galaxies in three broad stellar mass bins.
The central of these stellar mass bins was chosen to encompass as much as possible of the SFR difference seen in the {\it right} panel of Figure \ref{fig:sfrmass_relations} at $\sim 10^{10.5}$ $M_{\odot}$.
Uncertainties are estimated using the same bootstrapping methodology as mentioned earlier.
We find that at low- ($< 10^{10.1} M_{\odot}$) and high-stellar masses ($> 10^{10.8} M_{\odot}$) the SSFR of star-forming galaxies is largely independent of environment.
However, we see a $\approx$0.3 dex decrease in SSFR between the lowest and highest density environments for intermediate-stellar mass galaxies, significant at $\approx 4 \sigma$.

We further split galaxies into low- ($0.6 < z < 0.9$) and high-redshift subsamples ($0.9 < z < 1.3$) to see if any of these trends (or lack thereof) are driven by galaxies at a particular redshift (center and right panels of Figure \ref{fig:sfrmass_relations}).
The median redshifts of these subsamples are $z = 0.79$ and 1.05 respectively which are used as the ``fiducial'' redshift accordingly for applying the SFR correction described earlier.
In general we find that the behavior of the median SSFR versus environment in both the low- and high-redshift bins mirrors that of the full sample at 3.1$\sigma$ and 3.9$\sigma$ significance respectively, modulo a normalization offset of $0.1-0.2$ dex which is consistent with the redshift-evolution of the mean SFR-$M_*$ relation \citep[e.g.,][]{Speagle2014, Tomczak2016}.
Data points for measurements, as well as the full sample, are given in Table \ref{tab:ssfrs}.

\begin{table}

    \caption{Median SSFR-Density Relations of Star-Forming Galaxies}
    \label{tab:ssfrs}

    \begin{center}
    \begin{tabular}{rccc}

    \multicolumn{4}{c}{Full Sample ($0.6 < z < 1.3$)} \\[-1mm]
    \hline
    \hline

     & low & intermediate & high \\[-0.7mm]
    log($M_*$/$M_{\odot}$) & density & density & density \\[-0.5mm]
    \hline

    9.4$-$10.1  & $-9.00\pm0.02$ & $-8.95\pm0.04$ & $-9.05\pm0.05$ \\
    10.1$-$10.8 & $-9.22\pm0.02$ & $-9.28\pm0.04$ & $-9.48\pm0.06$ \\
    $>$10.8     & $-9.60\pm0.04$ & $-9.54\pm0.09$ & $-9.64\pm0.13$ \\

    \hline \\[-2mm]

    \multicolumn{4}{c}{Low-redshift ($0.6 < z < 0.9$)} \\[-1mm]
    \hline
    \hline

     & low & intermediate & high \\[-0.7mm]
    log($M_*$/$M_{\odot}$) & density & density & density \\[-0.5mm]
    \hline

    9.4$-$10.1  & $-9.08\pm0.03$ & $-9.02\pm0.09$ & $-9.12\pm0.07$ \\
    10.1$-$10.8 & $-9.32\pm0.02$ & $-9.36\pm0.04$ & $-9.61\pm0.09$ \\
    $>$10.8     & $-9.61\pm0.04$ & $-9.60\pm0.13$ & $-9.66\pm0.13$ \\

    \hline \\[-2mm]

    \multicolumn{4}{c}{High-redshift ($0.9 < z < 1.3$)} \\[-1mm]
    \hline
    \hline

     & low & intermediate & high \\[-0.7mm]
    log($M_*$/$M_{\odot}$) & density & density & density \\[-0.5mm]
    \hline

    9.4$-$10.1  & $-8.90\pm0.03$ & $-8.86\pm0.03$ & $-8.93\pm0.14$ \\
    10.1$-$10.8 & $-9.08\pm0.03$ & $-9.23\pm0.05$ & $-9.31\pm0.05$ \\
    $>$10.8     & $-9.54\pm0.05$ & $-9.47\pm0.11$ & $-9.39\pm0.22$ \\

    \hline
    \end{tabular}

    \end{center}
    Note: SSFRs are in units of log(yr$^{-1}$)
\end{table}

\section{A Simple Model to Contextualize Star-formation Histories}
\label{sec:simulation}

In this section we describe the framework and results of a toy model that we use to place qualitative and coarse quantitative constraints on the mean evolution of the star-formation histories (SFHs) of galaxies in the ORELSE data.
\citet{Wetzel2013} present a formula for the halo radius crossing time (assuming virial equilibrium), which is independent of host halo mass: $t_{\rm cross} \equiv R_{\rm vir} / V_{\rm vir} = 2.7 \times (1 + z)^{-3/2}$ Gyr.
The median redshifts of the high- and low-redshift subsamples considered in the previous section, $z = 1.05$ and $0.8$ respectively, correspond to $\approx$1.1 Gyr of cosmic time.
{\color{black}
Because this timescale is similar to $t_{\rm cross}$ at these redshifts an infalling galaxy at $z \approx 1.05$ would have time to transition from the outskirts to the core of a galaxy cluster/group over the time interval.
Therefore, we make the assumption that star-forming galaxies in all environments at $z \approx 0.8$ descended from star-forming galaxies in the field (i.e. low-density environments) at $z \approx 1.05$.
Thus, we will use the SSFRs between the two redshift subsamples as metrics to inform and test this toy model.
We will further discuss the effect of this assumption on the model at the end of this section.
}

We begin by simulating a sample of 10,000 galaxies at $z = 1.05$ (the median of our high-redshift bin), sampling stellar masses $>10^{8.7} M_{\odot}$ from the observed galaxy stellar mass function of star-forming galaxies \citep{Tomczak2014, Leja2015}.
This lower stellar mass limit is chosen so as to include all galaxies that can potentially grow to $M_* \geq 10^{9.4} M_{\odot}$ by $z = 0.8$, ensuring that all three stellar mass bins from earlier will be complete at this redshift.
Next, we assign SFRs to these galaxies that reproduce the observed SFR-$M_*$ relation at this redshift \citep{Tomczak2016} with $\pm$0.2 dex Gaussian scatter \citep{Speagle2014}.
In order to track the time-evolution of these simulated galaxies we make use of the Flexible Stellar Population Synthesis models \citep[FSPS:][]{Conroy2009, Conroy2010} using the {\tt python-fsps} code written by \citet{ForemanMackey2014}.
For these stellar population models we assume solar metallicity, a \citet{Chabrier2003} stellar IMF, no dust attenuation, and exponential SFHs of the form ${\rm SFR} \propto e^{-t/\tau}$.
We run several realizations of the model where for each realization all 10,000 galaxies are assigned the same value of $\tau$ from the set 
{\color{black}
[0.3, 0.5, 0.75, 1, 2, 4, $-$4] Gyr.
Although most of these values for $\tau$ correspond to declining SFHs we include one rising SFH ($\tau = -4$ Gyr) which some studies have found may be more representative of low-stellar mass galaxies at these redshifts \citep[e.g.,][]{Tomczak2016, Scoville2017}.
}

\begin{figure*}
    \includegraphics[width=2\columnwidth]{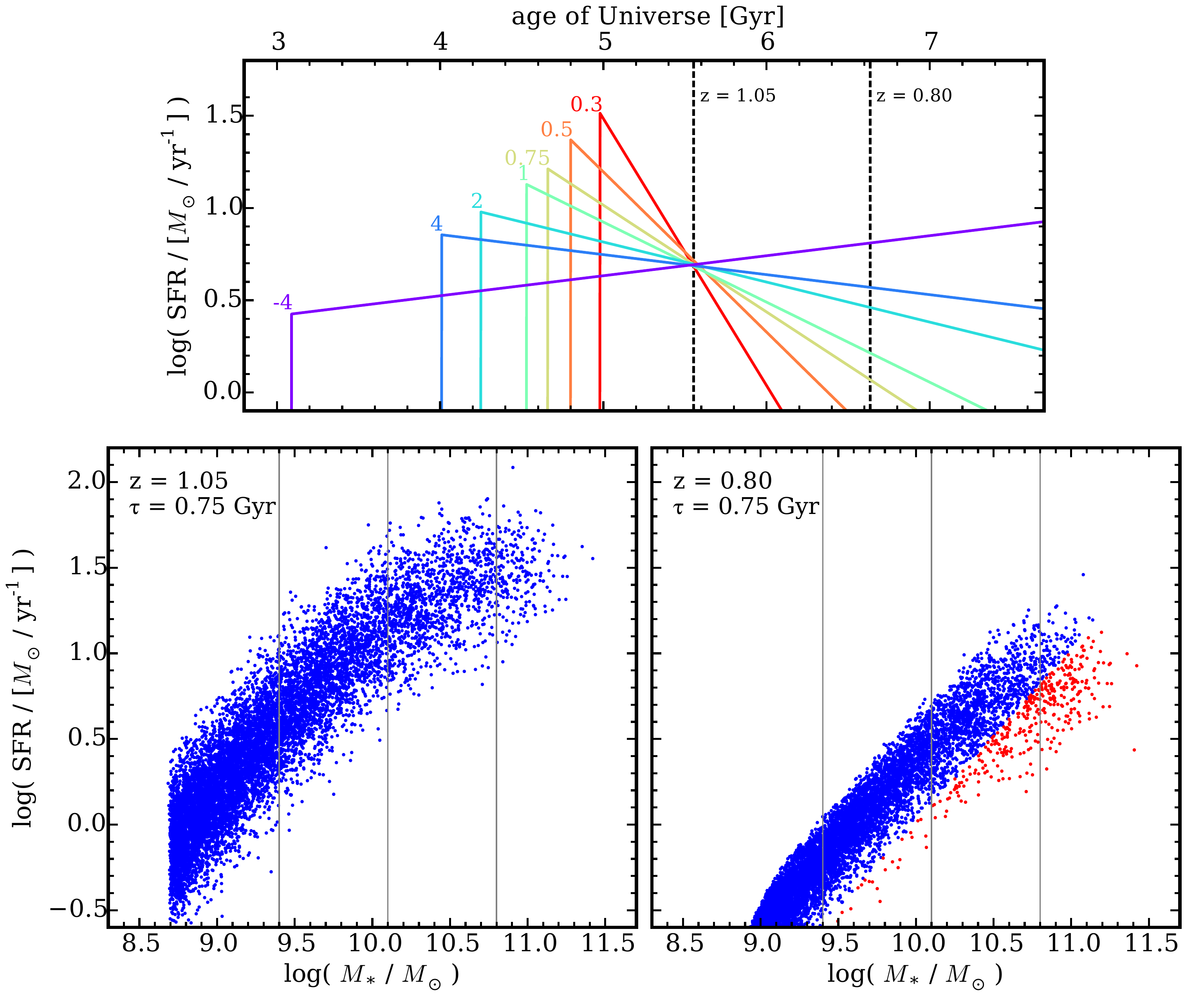}
    \caption{
    Schematic illustration of the model described in Section \ref{sec:simulation}.
    A sample of 10,000 simulated galaxies are generated and assigned stellar masses and SFRs that best reproduce the observed galaxy stellar mass function and SFR-$M_*$ relations at $z = 1.05$.
    We run several realizations of this sample, each time enforcing that all galaxies follow an 
    {\color{black}
    exponential SFH (${\rm SFR} \propto e^{-t/\tau}$) for each value of $\tau$ in [0.3, 0.5, 0.75, 1, 2, 4, -4] Gyr.
    }
    An example is shown in the top panel which plots these SFHs for a galaxy with $M_* = 10^{9.8} \: M_{\odot}$ and ${\rm SFR} = 5 \: M_{\odot}/{\rm yr}$ at $z = 1.05$.
    We then use the FSPS models to infer stellar mass growth and the evolution of the SED as a function of time.
    The two bottom panels show snapshots of these simulated galaxies in the SFR-$M_*$ plane at $z = 1.05$ and $0.8$ for the $\tau = 0.75$ Gyr realization.
    Vertical gray lines delineate the stellar mass bins considered in this study.
    Blue and red points in these panels correspond to galaxies classified as star-forming and quiescent respectively as defined by their rest-frame $(U - V)$ and $(V - J)$ colours (see Section \ref{sec:uvj}).
    }
    \label{fig:simulation_schematic}
\end{figure*}

Once a value of $\tau$ is chosen for a galaxy we check to make sure that the required formation time to produce its stellar mass is not greater than the age of the universe at $z = 1.05$, as this would be
{\color{black}
unphysical; such galaxies are not included in the simulation.
In practice the required formation time for the vast majority of galaxies is well within this constraint.
It only comes into play for the $\tau = 4$ Gyr and $-4$ Gyr realizations of the model where it predominantly affects galaxies that are $>2$$\sigma$ below the mean SFR-$M_*$ relation accounting for $<0.1$\%\ and 4.5\%\ of the full sample respectively.
We therefore neglect the effect of these galaxies given that they are so rare as to have no significant impact on the final result.
}
A schematic illustration of this model is shown in Figure \ref{fig:simulation_schematic}.

\begin{figure*}
    \includegraphics[width=2\columnwidth]{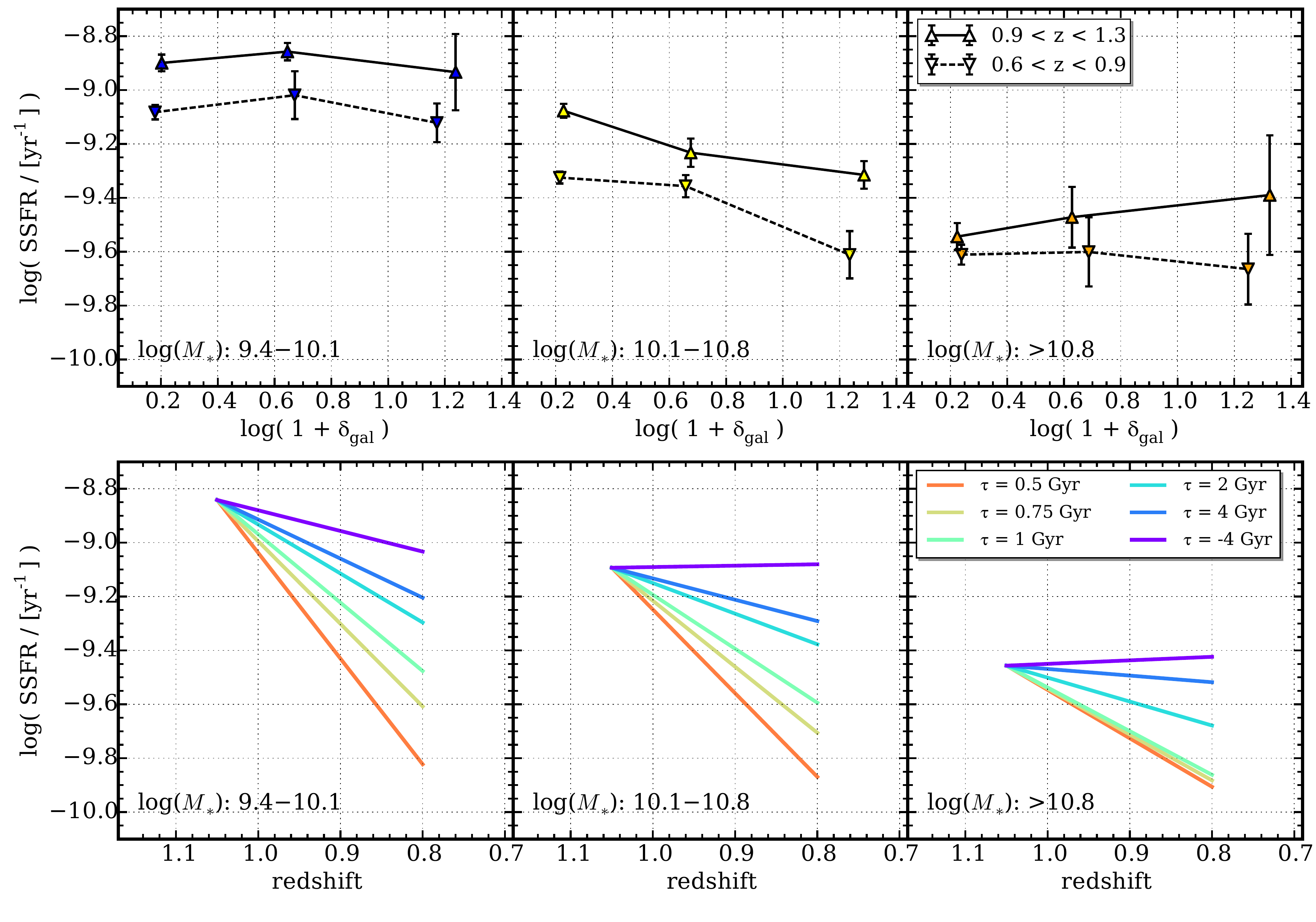}
    \caption{
    {\it Top row}: Median SSFRs of star-forming galaxies as a function of environment.
    Each panel corresponds to one of the stellar mass bins from Figure \ref{fig:ssfrdensity_relations} as indicated in the bottom-left corner.
    Upward-facing triangles with solid lines correspond to the high-redshift bin ($0.9 < z < 1.3$) whereas downward-facing triangles with dashed lines refer to the low-redshift bin ($0.6 < z < 0.9$).
    {\it Bottom row}: Median SSFRs of simulated star-forming galaxies in the same stellar mass bins as above at $z=1.05$ and $z=0.8$ (see Section \ref{sec:simulation} and Figure \ref{fig:simulation_schematic} for a details on the simulation).
    Galaxies with rest-frame colours consistent with quenched systems are removed from this portion of the analysis.
    Different colours correspond to different realizations of the simulation in which all galaxies follow an exponential star-formation history of the form ${\rm SFR} \propto e^{-t/\tau}$ adopting different values of $\tau$ as indicated in the legend.
    Tracks for $\tau = 0.3$ Gyr are not shown because all galaxies in this realization become classified as quiescent at $z = 0.8$.
    As can be seen, long$-\tau$ models ($\tau \gtrsim 2$ Gyr) are generally the best reproducing the observed drop in the median SSFR at nearly all stellar masses and environments at $z \sim 0.8$. 
    However, the larger SSFR decrement observed for intermediate-$M_*$ galaxies in high-density environments is better reproduced by the $\tau = 1$ Gyr model (assuming that $z=0.8$ star-forming galaxies in high-density environments descend from $z=1.05$ star-forming galaxies in low-density environments).
    }
    \label{fig:simulation_results}
\end{figure*}

\begin{figure*}
    \includegraphics[width=2\columnwidth]{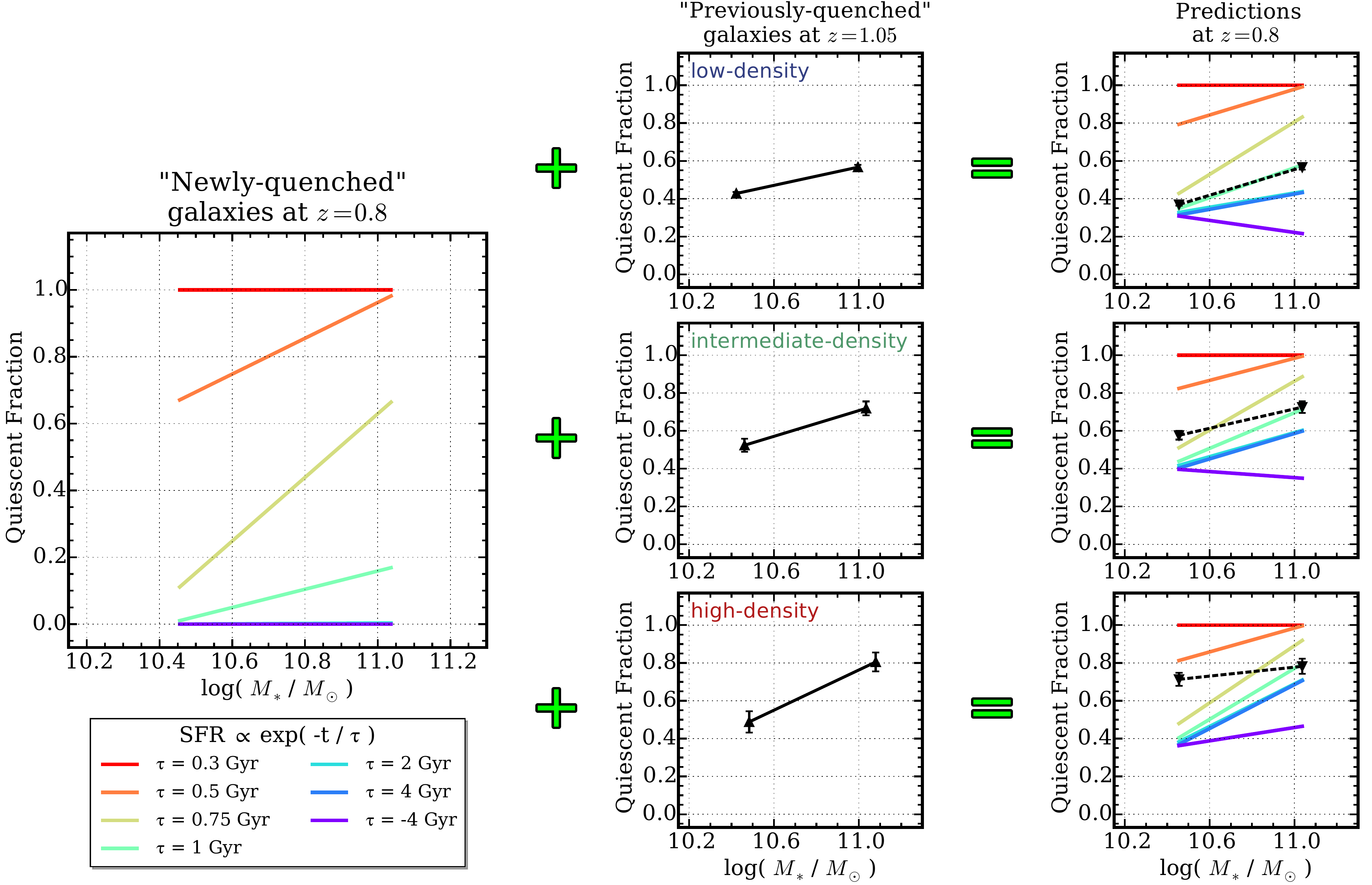}
    \caption{
    A schematic illustration of how quiescent fractions are calculated in the model presented in Section \ref{sec:simulation}.
    {\it Left}: 
    The fraction of simulated star-forming galaxies that quench over the time interval between $z=1.05-0.8$ ($\approx$1.1 Gyr) as a function of stellar mass for each $\tau$ realization of the model.
    We refer to these systems as ``newly-quenched'' galaxies.
    {\it Central column}: 
    Quiescent fractions measured from the ORELSE data
    {\color{black}
    in the high-redshift bin ($0.9 < z < 1.3$)
    }
    in each of the three environmental density bins explored in this study.
    Errorbars correspond to Poisson uncertainties.
    We use these measurements to infer the number of quenched galaxies that should exist at $z \approx 1.05$ in our simulation.
    We refer to these as ``previously-quenched'' galaxies and allow them to evolve coevally with the star-forming population.
    {\it Right column}: 
    Quiescent fraction predictions at $z=0.8$ for each realization of the model calculated by combining both groups of ``newly'' and ``previously'' quenched galaxies.
    Black points with dashed curves show the measured quiescent fractions at $0.6 < z < 0.9$.
    Note that effects of galaxy-galaxy merging and rejuvenation of star-formation are ignored in this model.
    }
    \label{fig:simulation_quiescent_fractions_schematic}
\end{figure*}

\subsection{Evolution of Median SSFR of Simulated Galaxies}

For each value of $\tau$ we calculate the median SSFR at $z = 1.05$ and $0.8$ in each of the three stellar mass bins explored in the previous section.
Results are shown in Figure \ref{fig:simulation_results}.
In this calculation we only include galaxies that are classified as star-forming based on the rest-frame $(U - V)$ and $(V - J)$ colours reported by the FSPS models and the same selection criteria that were applied to the ORELSE galaxies (see Section \ref{sec:uvj}).
{\color{black}
Realistic colour evolution of the FSPS models is demonstrated by \citet{Conroy2010} through a comparison with observations of local stellar populations.
Furthermore, because the rest-frame $(U-V)$ vs. $(V-J)$ colour selection method is generally insensitive to dust reddening we are confident that the classification of simulated galaxies is comparable to observed galaxies.
}
Note that the $\tau = 0.3$ Gyr realization is not plotted in Figure \ref{fig:simulation_results} because all star-forming galaxies at $z=1.05$ with this rapidly-declining SFH become classified as quenched by $z=0.8$.

For star-forming galaxies with high stellar masses ($>$$10^{10.8} M_{\odot}$) the median observed SSFR is independent of environment and displays a marginal decrease ($\lesssim 0.1$ dex) between the two redshift bins.
Note that there is a larger difference of $\approx$0.2 dex between the highest-density environments, but these measurements are poorly-constrained as they are based on only 27 galaxies and are statistically consistent with no evolution.
Comparing against the toy model, this observation is best reproduced with long e-folding timescales ($\tau \gtrsim 2$ Gyr).

Low-stellar mass ($10^{9.4} < M_* / M_{\odot} < 10^{10.1}$) star-forming galaxies show a slightly different picture.
Observationally, the median SSFRs of these galaxies show no statistically significant dependence on environmental density and are consistent with a uniform drop of $\approx$0.2 dex between the two redshift bins considered here.
{\color{black}
Curiously, this SSFR decrement is overproduced by all exponentially-declining SFHs and is rather best reproduced by a rising SFH with $\tau = -4$ Gyr.
Although SFRs are increasing in this realization of the model, the stellar masses of lower-$M_*$ galaxies are increasing more rapidly leading to a net decrease in the bulk SSFR.
}
At first this may seem somewhat paradoxical as the cosmic SFR density (SFRD) is decreasing at $z \sim 1$ \citep[e.g.,][]{Madau2014}; however, galaxies in this stellar mass window only account for roughly one third of the total SFRD, therefore it is possible for their mean SFR to be constant/increasing so long as other more massive galaxies' SFRs are decreasing to compensate.
Indeed, rising SFRs at low stellar masses are also seen in SFHs which are constructed by ``integrating'' along the mean SFR-$M_*$ relation \citep{Speagle2014, Tomczak2016}.

Arguably most interesting, however, are the intermediate-$M_*$ galaxies ($10^{10.1} < M_* / M_{\odot} < 10^{10.8}$) which show a statistically significant anticorrelation between median SSFR and environmental density in both redshift bins.
The magnitude of this difference is $\sim$0.2$-$0.3 dex between the lowest- and highest-density environments and is found at a $\geq 3 \sigma$ significance in both cases.
It is important to remember that this measurement is based on galaxies that were {\it pre-selected as star-forming} (see Section \ref{sec:uvj}) and is therefore not a reflection of a higher proportion of quenched galaxies in high-density environments.
This necessarily implies that environment exerts a non-negligible influence on the SFHs of these galaxies.
Within our model, under the premise that all galaxies in a given stellar mass bin have identical SFHs characterized be a single $\tau$, the median SSFR of intermediate-$M_*$ galaxies in low- and intermediate-density environments are best reproduced by the $\tau \approx 2$ Gyr realization, whereas in high-density environments this timescale is a factor of two lower.

However, the exact nature of an environmental influence can be complex and governed by multiple non-linear and/or codependent mechanisms such as galaxy-galaxy mergers, tidal interactions, bursts of star-formation, etc.
For instance, using cosmological $N$-body simulations and subhalo abundance matching \citet{Wetzel2013} explored quenching pathways of galaxies as a function of environment and stellar mass.
They find that in the most massive host halos ($\geq$$10^{14}$ $M_{\odot}$), intermediate-$M_*$ quiescent galaxies predominantly experience their ``quenching event'' within their current host halo ($\approx 50$\%), although a reasonable proportion experience preprocessing ($\approx 30$\%) where quenching occurs in a galaxy group environment prior to infall into a more massive halo.
These different pathways are likely to exert different quenching mechanisms on galaxies.
In the remaining sections we attempt to investigate some of these pathways.

\begin{figure*}
    \includegraphics[width=2\columnwidth]{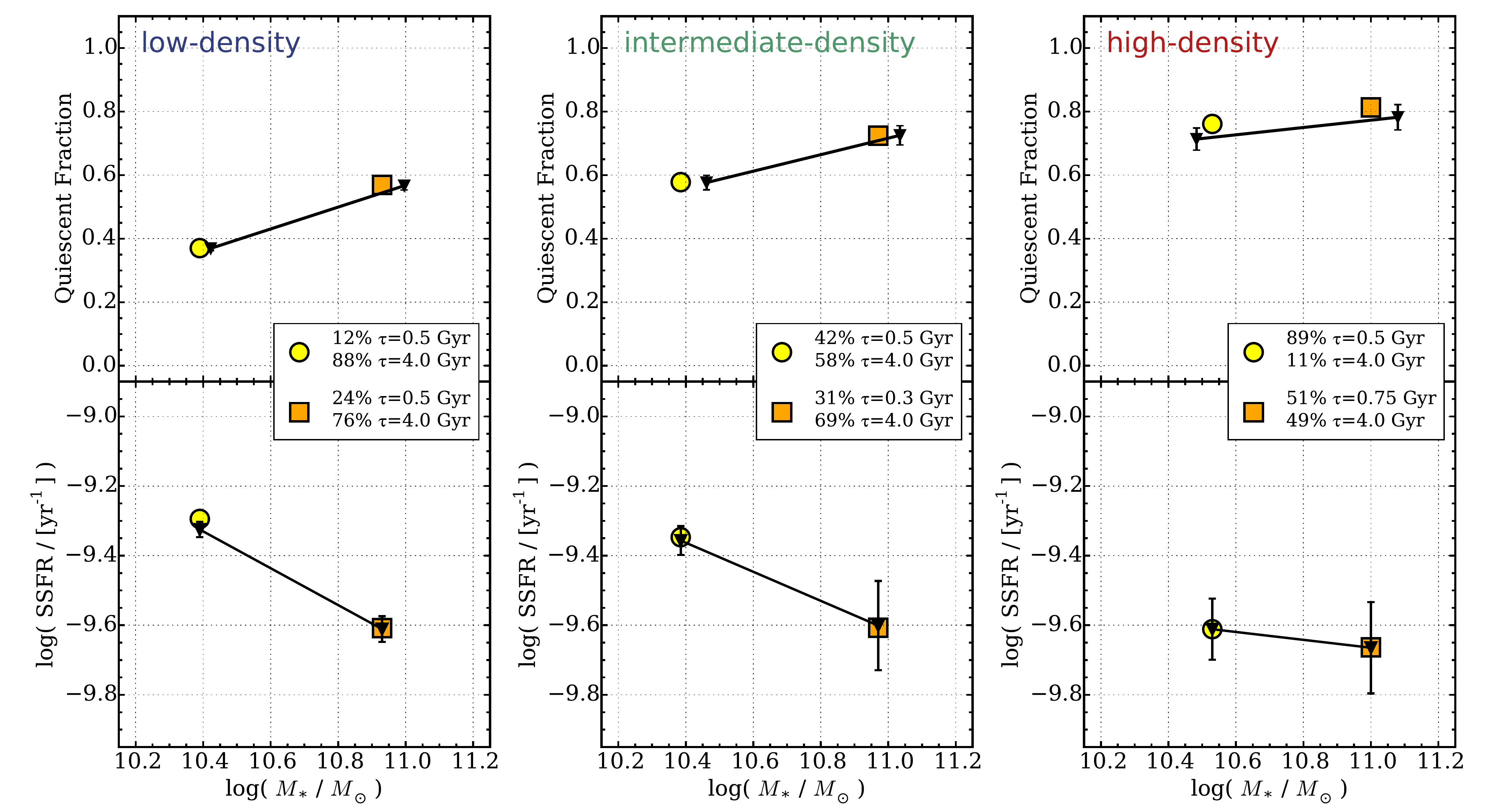}
    \caption{
    Quiescent fractions ({\it top row}) and SSFRs of star-forming galaxies ({\it bottom row}) for each of the three environmental density bins in the two highest stellar mass bins considered in this study.
    Black triangles correspond to measurements from the ORELSE data at $0.6 < z < 0.9$ with errorbars corresponding to Poisson uncertainties and bootstrapped errors for quiescent fractions and SSFRs respectively.
    Results from the model discussed in Section \ref{sec:simulation} are also shown.
    Briefly, the model treats all galaxies in a given environment-stellar mass bin as a linear combination of two subpopulations of galaxies following long$-\tau$ and short$-\tau$ exponential star-formation histories (${\rm SFR} \propto e^{-t/\tau}$), yielding simultaneous predictions for the quiescent fraction and SSFR.
    Yellow circles and orange squares show results from this model for each environment-stellar mass bin with best-fit values for the long$-\tau$ and short$-\tau$ timescales and their respective weights indicated in the legends.
    In general there is a transition between a preference for long$-\tau$ SFHs at low environmental density to short$-\tau$ SFHs at high environmental density.
    This transition is most prominent for intermediate-$M_*$ galaxies ($10^{10.1} < M_* / M_{\odot} < 10^{10.8}$) and is even apparent between the low- and intermediate-density bins, potentially suggesting environmental preprocessing.
    }
    \label{fig:simulation_fits}
\end{figure*}

\subsection{Quiescent Fractions of Simulated Galaxies}

Galaxies do not remain star-forming forever.
Indeed, a galaxy with $\tau = 0.1$ Gyr is expected to have rest-frame colours consistent with quenched galaxies after only $\approx$0.4 Gyr \citep{Moutard2016}.
As mentioned earlier, we track the evolution of the rest-frame $(U - V)$ and $(V - J)$ colours of simulated galaxies for the purpose of identifying at what times they would be classified as ``star-forming'' versus ``quiescent''.
This naturally yields predictions for the quiescent fraction ($f_{\rm q}$) which can be compared to observations.
In Figure \ref{fig:simulation_quiescent_fractions_schematic} we illustrate how we estimate quiescent fractions in our toy model in the same stellar mass and environmental density bins as in Figures \ref{fig:ssfrdensity_relations} and \ref{fig:simulation_results}.
Because the typical stellar mass completeness limit for maximally red (e.g. quenched) galaxies in our imaging data at the LSS redshifts is around $10^{10} M_{\odot}$ \citep[see Section 3.1 of][]{Tomczak2017} we restrict the remainder of our analysis to the two higher stellar mass bins ($M_* > 10^{10.1} M_{\odot}$) where we are complete for all types of galaxies.

Note that up to this point the model has only simulated the subpopulation of star-forming galaxies at $z=1.05$; therefore, the resulting quiescent fractions at $z=0.8$ will only include galaxies that quench their star-formation over the $\approx 1.1$ Gyr time interval between these redshifts.
We refer to these galaxies as ``newly-quenched'' and plot their fraction as a function of stellar mass in the leftmost panel of Figure \ref{fig:simulation_quiescent_fractions_schematic}.
As can be seen, in the shortest $\tau$ realization (0.3 Gyr) all galaxies which begin as star-forming at $z=1.05$ quench by $z=0.8$ whereas in the longest $\tau$ realization (4 Gyr) 
{\color{black}
and the rising SFH
}
no star-forming galaxies quench.

To make these quiescent fractions more representative of reality we add in galaxies that are quenched at the beginning of the simulation.
These ``previously-quenched'' galaxies are added so as to reproduce the observed quiescent fractions as a function of stellar mass at $z=1.05$.
We follow a similar procedure as described in Section 3.4 of \citet{Tomczak2017} to directly measure the quiescent fraction of galaxies.
This is done by counting the numbers of galaxies in the full photometric + spectroscopic ORELSE dataset in the redshift, stellar mass, and environmental density bins considered here.
The central column of Figure \ref{fig:simulation_quiescent_fractions_schematic} shows the measured quiescent fractions at $z=1.05$ which we use to infer the number of quenched galaxies that are not accounted for at the beginning of the model.
A more thorough investigation of the quiescent fraction as a function of redshift, stellar mass, and environment will be presented in a future paper (Lemaux et al. in preparation).
Assuming these galaxies evolve statically, alongside the simulated star-forming population, we combine them with the ``newly-quenched'' galaxies at $z=0.8$ to produce predictions of quiescent fractions for each $\tau$ realization of the model (shown in the rightmost column of Figure \ref{fig:simulation_quiescent_fractions_schematic}).

It is important to note that effects of galaxy-galaxy merging are not included in this model.
Simulations \citep[e.g.,][]{Hopkins2010, RodriguezGomez2015} and observations \citep[e.g.,][]{Lin2008, Lotz2011, Man2016} generally agree that the major merger rate per galaxy at $z \approx 1$ is around 0.1 Gyr$^{-1}$.
This estimate, however, applies to the general ``field'' galaxy population and could be a factor 2$\times$ greater in denser environments \citep{Lin2010}.
This could potentially introduce additional systematic uncertainties when comparing to observations, specifically if merging preferentially ``destroys'' one galaxy type over another (i.e. star-forming versus quiescent).
The primary challenge in incorporating mergers into the toy model is in deciding the what happens to the resulting, post-merger galaxy.
Simulations have shown that mergers increase the efficiency of star-formation within the associated galaxies \citep[e.g.,][]{Cox2008, Sparre2016}.
However, whether or not this process accelerates quenching or prolongs the star-forming state of a galaxy is unclear, especially over the relatively short ($\approx 1$ Gyr) timescale considered in this work.
Recent studies have found that mergers can, but do not necessarily lead to quenching, arguing that post-merger quenching only occurs in cases where strong AGN feedback processes are in effect \citep{Pontzen2017, Sparre2017}.
Even in cases where post-merger quenching occurs the timescale is poorly constrained, varying widely between 250 Myr to 3 Gyr.
Therefore, in the interest of maintaining simplicity within the toy model presented here, we have decided to not implement a strategy for galaxy-galaxy merging.

\subsection{Fitting the Model to Observations}

As described in the preceding sections, each realization of the model produces a distinct set of predictions for the evolution of the SSFR and quiescent fraction of galaxies.
In Figures \ref{fig:simulation_results} and \ref{fig:simulation_quiescent_fractions_schematic} we show these results binned by stellar mass and environment.
Upon careful examination it is clear that no single $\tau$ realization can simultaneously reproduce both the observed median SSFR of star-forming galaxies and the quiescent fraction at $z=0.8$ for any environment-stellar mass bin.
This is perhaps not surprising because it would be an oversimplification to assume that all galaxies, even at fixed environment and stellar mass, should follow a similar star-formation history.

Therefore, to add one more layer to the model without overcomplicating it we assume that star-forming galaxies in a given environment-stellar mass bin can be described as a linear combination of two exponentially-declining SFHs.
For brevity we refer to these two subsets as ``long$-\tau$'' and ``short$-\tau$'' galaxies.
In all cases we fix the timescale for long$-\tau$ galaxies to $\tau = 4$ Gyr.
This rate is broadly consistent with the bulk evolution of the cosmic star-formation rate density at $z \approx 1$ \citep{Madau2014}, therefore making it a reasonable assumption for a class of ``normal'' star-forming galaxies.
Nevertheless, our final results are unchanged if we allow the long$-\tau$ value to be a free parameter.
Correspondingly, the short$-\tau$ galaxies can be thought of as representing those star-forming galaxies that are in the process of quenching.
{\color{black}
For each long$-\tau$/short$-\tau$ pair in each environment-stellar mass bin we identify the linear combination that minimizes the residual $\chi^2$ sum as defined as:

$$
\mathrm{\chi^2 \: \equiv \; \left( \frac{\Delta log(SSFR)}{\sigma_{SSFR}} \right)^2 \; + \; \left( \frac{\Delta log({\it f}_q)}{\sigma_{{\it f}_q}} \right)^2}
$$ 
$$
\mathrm{\Delta log(SSFR) \; = \; log \left( \frac{SSFR_{observed}}{SSFR_{model}} \right)}
$$
$$
\mathrm{\Delta log({\it f}_q) \; = \; log \left( \frac{{\it f}_{q, observed}}{{\it f}_{q, model}} \right)}
$$
\\

\noindent
where $\sigma_{SSFR}$ and $\sigma_{{\it f}_q}$ represent the uncertainties on the observed log(SSFR) and log($f_{\rm q}$) respectively.
}

In Figure \ref{fig:simulation_fits} we show the resulting best-fit linear combinations for each of the six bins examined in this portion of the analysis (three bins of environment and two bins of stellar mass).
As can be seen, a simple linear combination of two $\tau$ models is capable of accurately reproducing the observations.
The largest residuals between the observations and model for the median SSFR and quiescent fraction are 4\%\ and 5\%\ respectively.
Even though the short$-\tau$ timescale is allowed to vary, intermediate-$M_*$ galaxies ($10^{10.1} < M_* / M_{\odot} < 10^{10.8}$) at all environmental densities always prefer $\tau = 0.5$ Gyr for the short$-\tau$ galaxies.
Furthermore, the relative weights of short$-\tau$ versus long$-\tau$ galaxies varies dramatically as a function of environment shifting from (12\%, 88\%) at low-density to (42\%, 58\%) at intermediate-density to (89\%, 11\%) at high-density.
At face value these numbers suggest that environmental quenching (1) is highly efficient for intermediate-$M_*$ galaxies and (2) begins to appear at intermediate densities supporting a ``preprocessing'' scenario in which some galaxies quench as satellites of halos less massive than the halo they are observed in \citep[e.g.,][]{Wetzel2013, Hou2014, Haines2015}.
This picture is similar, though slightly different for galaxies in the high-$M_*$ bin ($M_* / M_{\odot} > 10^{10.8}$).
The preferred short$-\tau$ timescale changes as a function of environment and, while not always best fit by $\tau = 0.5$ Gyr, is still always rapid ($\leq$0.75 Gyr).
The variation in the relative weights of short$-\tau$ versus long$-\tau$ galaxies is less extreme than in the intermediate-$M_*$ bin, but still has similar behavior, shifting from (24\%, 76\%) at low-density to (31\%, 69\%) at intermediate-density to (51\%, 49\%) at high-density.
These numbers suggest that at high stellar masses environmental quenching is still relevant, though to a lesser degree than for less massive galaxies,
{\color{black}
and that ``preprocessing'' in intermediate-density environments, if present, is much weaker.
}
This is broadly consistent with the findings of \citet{Wetzel2013} where the authors show that at high stellar masses, galaxies are most likely to have quenched either prior to infall into a cluster or within the cluster halo where it is observed whereas less massive galaxies are most like to quench either via group preprocessing or within the cluster halo where it is observed (see their Figure 10).

Results from the toy model are also consistent with environmental studies of recently-quenched galaxies, commonly referred to as ``post-starburst'' galaxies.
The preference for shorter star-forming timescales towards higher environmental densities necessarily implies a higher incidence of post-starburst galaxies in these denser environments.
In the local universe post-starburst galaxies have been shown to predominantly reside in low density environments \citep[e.g.,][]{Zabludoff1996, Hogg2006, Wilkinson2017, Pawlik2018}; however, this trend reverses at higher redshifts ($z \gtrsim 0.4$) wherein post-starbursts are more commonly found in higher-density environments \citep[e.g.,][]{Dressler1999, Poggianti2009, Yan2009, Muzzin2012, Wu2014, Lemaux2017}.
Two of these studies \citep{Wu2014, Lemaux2017} are based on specific large scale structures included in the present work.

{\color{black}
It is important to reiterate that this model assumes that {\it star-forming} galaxies in all environments at low-$z$ descended from {\it star-forming} galaxies in the field (i.e. low-density) at high-$z$.
This is based on the circumstance that the length of the time interval between $z = 1.05 - 0.8$ is comparable to the halo radius crossing time which, for example, implies that a galaxy on the outskirts would have time to migrate to the core of a cluster.
A relevant caveat to note is that some SF galaxies could have migrated from other environments (e.g. from intermediate- to high-density) or remained in the same environment.
However, if we also assume that SF galaxies that spend the longest time in dense environments will be the first to quench then it follows that the surviving SF galaxies at low-$z$ would most likely be those that were accreted most recently, i.e., originated from low-density environments (for further discussion see Lemaux et al. in preparation).
Furthermore, this is not a concern for galaxies in the low- or high-$M_*$ bins because our measurements suggest that SSFR is independent of environment for these galaxies (see Figure \ref{fig:ssfrdensity_relations}).
On the other hand, for intermediate-$M_*$ galaxies this could impose a small systematic bias.
If some SF galaxies in the low-$z$ bin originated from denser environments at high-$z$ it would lower the initial SFR.
Consequently, this would lessen the SSFR decrement between the two redshift bins.
In the context of the long$-\tau$/short$-\tau$ combination this would imply either (1) a larger value for the short$-\tau$ or (2) a smaller proportion of short$-\tau$ galaxies.
However, because of the quiescent fraction acts as an additional constraint we suspect that the impact would not greatly alter the results.
}

\section{Summary and Conclusions}
\label{sec:conclusions}

In this paper we examine the relations between star-formation rate, stellar mass, and environment for 5770 spectroscopically-selected galaxies at $z \sim 0.9$.
Galaxies are drawn from the ORELSE and VVDS spectroscopic surveys which predominantly probe high- and low-density environments, respectively.
The ORELSE sample is composed of galaxies from 12 massive ($M_{\rm vir} > 10^{14.5} M_{\odot}$) large scale structures, many composed of multiple galaxy groups and/or clusters, in the redshift range $0.7 \leq z \leq 1.26$.
Each field has broadband photometric observations in $\geq$9 filters which are used to estimate stellar masses as well as rest-frame ($U - V$) and ($V - J$) colours.
Star-formation rates are estimated by adding contributions from obscured and unobscured star-forming regions as traced by IR and UV emission respectively.

Environment is measured using a Monte Carlo Voronoi tessellation technique which is sensitive to the local environmental density of galaxies.
Using this metric we separate galaxies into three bins of environment: low-, intermediate-, and high-density.
Among these subsamples we calculate the median SFR$_{\rm UV+IR}$ as a function of stellar mass.
At all $M_* \gtrsim 10^{9.5} M_{\odot}$ we see a strong anticorrelation between SFR and environmental density at fixed stellar mass.
However, when considering only actively star-forming galaxies (i.e. excluding galaxies classified as quiescent) this difference largely vanishes, indicating that this effect is predominantly a reflection of higher-density environments at $z \sim 0.9$ hosting a higher proportion of quenched galaxies relative to the field.

Nevertheless, {\it even among star-forming galaxies} there still remains a statistically significant anticorrelation between SFR and density suggesting that environmental processes that regulate star-formation are at work.
Interestingly, this residual SFR-density anticorrelation appears to only affect galaxies with intermediate stellar masses ($10^{10.1} < M_* / M_{\odot} < 10^{10.8}$) whereas lower- and higher-$M_*$ galaxies show no dependence.
We find that the median SSFR for intermediate-$M_*$ galaxies in the highest-density environments is a factor of 2$\times$ lower than that of their counterparts in the field at a statistical significance of $\approx 4$$\sigma$.
These environmental dependencies and the statistical significance of this offset remain virtually unchanged when splitting galaxies into low-redshift ($0.6 < z < 0.9$) and high-redshift ($0.9 < z < 1.3$) subsamples.

To investigate this offset further we create a simple model with the general concept of evolving a sample of simulated star-forming galaxies from the high-redshift to low-redshift bins mentioned above.
Simulated galaxies are assigned stellar masses and star-formation rates that match the observed galaxy stellar mass function and SFR-$M_*$ relation of star-forming galaxies for the high-redshift bin.
We use the FSPS model library \citep{Conroy2009, Conroy2010} to track the evolution of galaxy SFRs, stellar masses, and rest-frame colours in a realistic manner.
Adopting exponential star-formation histories (${\rm SFR} \propto e^{-t/\tau}$) we evolve this simulated sample to the low-redshift bin assuming different values of the timescale $\tau$ and examine the evolution of the median SSFR and quiescent fraction of galaxies.
We assume that star-forming galaxies can be approximated as a linear combination of two SFR timescales: a long$-\tau$ subset representing ``normal'' star-forming galaxies and a short$-\tau$ subset representing those galaxies that are on a quenching pathway.
The value for the long$-\tau$ group is fixed at $\tau = 4$ Gyr which is roughly equivalent to the evolution of the bulk cosmic star-formation rate density at $z = 1$ \citep{Madau2014}.
For the sake of simplicity we do not include galaxy-galaxy mergers as a feature in this toy model given the complex nature of how star-formation is affected in the aftermath of a merger \citep{Cox2008, Sparre2016, Pontzen2017, Sparre2017}.
We fit this compound model to each environment-stellar mass bin independently.

In general we find that our observations at low environmental densities tend to prefer versions of the model that are more heavily weighted by the long$-\tau$ population whereas at high environmental densities they tend to prefer versions more heavily weighted by the short$-\tau$ population.
This indicates that environmental quenching is going on at $z \approx 0.9$, i.e. dense galactic environments promote quenching of star-formation at this redshift, consistent with findings from recent studies \citep{Balogh2016, Nantais2017}.
This result is also consistent with studies that show an increasing fraction of post-starburst galaxies in higher-density environments at these redshifts \citep[e.g.,][]{Dressler1999, Poggianti2009, Yan2009, Muzzin2012, Wu2014, Lemaux2017}.
Furthermore, this transition between long$-\tau$ to short$-\tau$ dominance appears to occur at intermediate-densities for intermediate-$M_*$ galaxies ($10^{10.1} < M_* / M_{\odot} < 10^{10.8}$) suggesting that quenching via ``preprocessing'' is a relevant mechanism for these galaxies.
This finding is also consistent with recent studies in the low-redshift universe \citep{Wetzel2013, Hou2014, Haines2015}.

\section*{Acknowledgements}

This material is based upon work supported by the National Aeronautics and Space Administration under NASA Grant Number NNX15AK92G.
Part of the work presented herein is supported by the National Science Foundation under Grant No. 1411943.
This research made use of Astropy, a community-developed core Python package for Astronomy (Astropy Collaboration, 2013).
A.R.T. would also like to thank Chris Fassnacht for aiding in the reduction of photometric data through helpful discussions and provision of well-documented code.
PFW acknowledges funding through the H2020 ERC Consolidator Grant 683184.
Work presented here is based in part on data collected at Subaru Telescope as well as archival data obtained from the SMOKA, which is operated by the Astronomy Data Center, National Astronomical Observatory of Japan.
This work is based in part on observations made with the Large Format Camera mounted on the 200-inch Hale Telescope at Palomar Observatory, owned and operated by the California Institute of Technology.
A subset of observations were obtained with WIRCam, a joint project of CFHT, Taiwan, Korea, Canada, France, at the Canada-France-Hawaii Telescope (CFHT) which is operated by the National Research Council (NRC) of Canada, the Institut National des Sciences de l'Univers of the Centre National de la Recherche Scientifique of France, and the University of Hawaii.
UKIRT is supported by NASA and operated under an agreement among the University of Hawaii, the University of Arizona, and Lockheed Martin Advanced Technology Center; operations are enabled through the cooperation of the East Asian Observatory.
When the data reported here were acquired, UKIRT was operated by the Joint Astronomy Centre on behalf of the Science and Technology Facilities Council of the U.K. 
This work is based in part on observations made with the Spitzer Space Telescope, which is operated by the Jet Propulsion Laboratory, California Institute of Technology under a contract with NASA.
Spectroscopic observations used in the work presented here were obtained at the W.M. Keck Observatory, which is operated as a scientific partnership among the California Institute of Technology, the University of California and the National Aeronautics and Space Administration.
The Observatory was made possible by the generous financial support of the W.M. Keck Foundation.
We wish to recognize and acknowledge the very significant cultural role and reverence that the summit of Mauna Kea has always had within the indigenous Hawaiian community.
We are most fortunate to have the opportunity to conduct observations from this mountain.

\appendix

\section{Photometric Data}
\label{appendix}

% ~~~~~~~~~~~~~~~~~~~~~~~~~~~~~~~~~~~~~~~~~~~~~~~~~~~~~~~~~~~~~
% ~~~~~~~~~~~~~~~~~~~~~~~~~~~~~ TABLE ~~~~~~~~~~~~~~~~~~~~~~~~~~
% ~~~~~~~~~~~~~~~~~~~~~~~~~~~~~~~~~~~~~~~~~~~~~~~~~~~~~~~~~~~~~

\begin{table*}
        \begin{center}
        \caption{Photometry}
        \label{tab:photometry1}

        {\vskip 1mm}
        \begin{tabular}{c @{\hskip 15mm} c}

                \begin{tabular}{llll}

                \hline \\[-3.3mm]  

                Filter & Telescope & Instrument & Depth$^a$ \\[0mm]

                \hline \\[-3mm]  
                SC1604 \\[-1mm]
                \hline \\[-5mm]  
                \hline \\[-2mm]

                $B$    &   Subaru   &   Suprime-Cam   &   26.6   \\
                $V$    &   Subaru   &   Suprime-Cam   &   26.1   \\
                $R_C$  &   Subaru   &   Suprime-Cam   &   26.0   \\
                $I_C$  &   Subaru   &   Suprime-Cam   &   25.1   \\
                $Z_+$  &   Subaru   &   Suprime-Cam   &   24.6   \\
                $r'$   &   Palomar   &   LFC   &   24.2   \\
                $i'$   &   Palomar   &   LFC   &   23.6   \\
                $z'$   &   Palomar   &   LFC   &   23.1   \\
                $J$   &   UKIRT   &   WFCAM   &   22.1   \\
                $K$   &   UKIRT   &   WFCAM   &   21.9   \\
                $[3.6]$  &   {\it Spitzer}  &   IRAC   &   24.7   \\
                $[4.5]$  &   {\it Spitzer}  &   IRAC   &   24.3   \\
                $[5.8]$  &   {\it Spitzer}  &   IRAC   &   22.7   \\
                $[8.0]$  &   {\it Spitzer}  &   IRAC   &   22.6   \\[1mm]

                \hline \\[-3mm]
                RXJ1716 \\[-1mm]
                \hline \\[-5mm]  
                \hline \\[-2mm]

                $B$    &   Subaru   &   Suprime-Cam   &   25.9   \\
                $V$    &   Subaru   &   Suprime-Cam   &   26.6   \\
                $R_C$  &   Subaru   &   Suprime-Cam   &   26.2   \\
                $I_+$  &   Subaru   &   Suprime-Cam   &   25.4   \\
                $Z_+$  &   Subaru   &   Suprime-Cam   &   24.7   \\
                $J$    &   CFHT   &   WIRCam   &   21.3   \\
                $K_s$  &   CFHT   &   WIRCam   &   21.7   \\
                $[3.6]$  &   {\it Spitzer}  &   IRAC   &   24.6   \\
                $[4.5]$  &   {\it Spitzer}  &   IRAC   &   24.1   \\
                $[5.8]$  &   {\it Spitzer}  &   IRAC   &   22.4   \\
                $[8.0]$  &   {\it Spitzer}  &   IRAC   &   22.3   \\[1mm]

                \hline \\[-3mm]  
                SG0023 \\[-1mm]
                \hline \\[-5mm]  
                \hline \\[-2mm]

                $B$      &   Subaru         &   Suprime-Cam   &   26.4   \\
                $V$      &   Subaru         &   Suprime-Cam   &   26.2   \\
                $R_+$    &   Subaru         &   Suprime-Cam   &   25.3   \\
                $I_+$    &   Subaru         &   Suprime-Cam   &   25.2   \\
                $r'$     &   Palomar        &   LFC           &   25.7   \\
                $i'$     &   Palomar        &   LFC           &   25.2   \\
                $z'$     &   Palomar        &   LFC           &   23.8   \\
                $J$      &   UKIRT          &   WFCAM         &   21.6   \\
                $K$      &   UKIRT          &   WFCAM         &   21.6   \\
                $[3.6]$  &   {\it Spitzer}  &   IRAC          &   24.0   \\
                $[4.5]$  &   {\it Spitzer}  &   IRAC          &   23.8   \\[1mm]

                \hline \\[-3mm]
                RCS0224 \\[-1mm]
                \hline \\[-5mm]  
                \hline \\[-2mm]

                $B$    &   Subaru   &   Suprime-Cam   &   26.2   \\
                $V$    &   Subaru   &   Suprime-Cam   &   26.0   \\
                $R_+$    &   Subaru   &   Suprime-Cam   &   25.9   \\
                $I_+$    &   Subaru   &   Suprime-Cam   &   25.5   \\
                $Z_+$    &   Subaru   &   Suprime-Cam   &   24.9   \\
                $J$   &   UKIRT   &   WFCAM   &   21.2   \\
                $K$   &   UKIRT   &   WFCAM   &   21.4   \\
                $[3.6]$  &   {\it Spitzer}  &   IRAC   &   24.0   \\
                $[4.5]$  &   {\it Spitzer}  &   IRAC   &   23.6   \\[1mm]

                \hline \\[-3mm]

                \end{tabular}

                &

                \begin{tabular}{llll}

                \hline \\[-3mm]  

                Filter & Telescope & Instrument & Depth$^a$ \\[0mm]

                \hline \\[-3mm]  
                SC0849 \\[-1mm]
                \hline \\[-5mm]  
                \hline \\[-2mm]

                $B$    &   Subaru   &   Suprime-Cam   &   26.4   \\
                $V$    &   Subaru   &   Suprime-Cam   &   26.5   \\
                $R_C$  &   Subaru   &   Suprime-Cam   &   26.2   \\
                $I_+$  &   Subaru   &   Suprime-Cam   &   25.5   \\
                $Z_+$  &   Subaru   &   Suprime-Cam   &   25.1   \\
                $Z_R$  &   Subaru   &   Suprime-Cam   &   23.5   \\
                $r'$   &   Palomar   &   LFC   &   24.7   \\
                $i'$   &   Palomar   &   LFC   &   24.4   \\
                $z'$   &   Palomar   &   LFC   &   23.3   \\
                $NB711$  &   Subaru   &   Suprime-Cam   &   23.7   \\
                $NB816$  &   Subaru   &   Suprime-Cam   &   25.9   \\
                $J$   &   UKIRT   &   WFCAM   &   21.8   \\
                $K$   &   UKIRT   &   WFCAM   &   21.6   \\
                $[3.6]$  &   {\it Spitzer}  &   IRAC   &   24.8   \\
                $[4.5]$  &   {\it Spitzer}  &   IRAC   &   24.3   \\[1mm]

                \hline \\[-3mm]
                CL1429 \\[-1mm]
                \hline \\[-5mm]  
                \hline \\[-2mm]

                $B$    &   Subaru   &   Suprime-Cam   &   26.7   \\
                $V$    &   Subaru   &   Suprime-Cam   &   26.2   \\
                $r'$     &   Palomar   &   LFC   &   24.2   \\
                $i'$     &   Palomar   &   LFC   &   23.5   \\
                $z'$     &   Palomar   &   LFC   &   22.7   \\
                $Y$    &   Subaru  &   Suprime-Cam   &   23.2   \\
                $J$    &   UKIRT   &   WFCAM   &   21.9   \\
                $K$    &   UKIRT   &   WFCAM   &   21.7   \\
                $[3.6]$  &   {\it Spitzer}  &   IRAC   &   23.1   \\
                $[4.5]$  &   {\it Spitzer}  &   IRAC   &   23.1   \\[1mm]

                % aka nep5281
                \hline \\[-3mm]
                RXJ1821 \\[-1mm]
                \hline \\[-5mm]  
                \hline \\[-2mm]

                $B$    &   Subaru   &   Suprime-Cam   &   26.0   \\
                $V$    &   Subaru   &   Suprime-Cam   &   26.0   \\
                $r'$   &   Palomar   &   LFC   &   24.4   \\
                $i'$   &   Palomar   &   LFC   &   24.3   \\
                $z'$   &   Palomar   &   LFC   &   23.3   \\
                $Y$    &   Subaru   &   Suprime-Cam   &   23.4   \\
                $J$     &   CFHT   &   WIRCam   &   21.4   \\
                $K_s$   &   CFHT   &   WIRCam   &   21.7   \\
                $[3.6]$  &   {\it Spitzer}  &   IRAC   &   23.9   \\
                $[4.5]$  &   {\it Spitzer}  &   IRAC   &   23.8   \\[1mm]

                \hline \\[-3mm]
                SC0910 \\[-1mm]
                \hline \\[-5mm]  
                \hline \\[-2mm]

                $B$    &   Subaru   &   Suprime-Cam   &   24.4   \\
                $V$    &   Subaru   &   Suprime-Cam   &   25.6   \\
                $R_C$  &   Subaru   &   Suprime-Cam   &   26.4   \\
                $I_+$  &   Subaru   &   Suprime-Cam   &   25.8   \\
                $Z_+$  &   Subaru   &   Suprime-Cam   &   24.8   \\
                $J$   &   UKIRT   &   WFCAM   &   22.1   \\
                $K$   &   UKIRT   &   WFCAM   &   21.7   \\
                $[3.6]$  &   {\it Spitzer}  &   IRAC   &   23.2   \\
                $[4.5]$  &   {\it Spitzer}  &   IRAC   &   23.2   \\[1mm]

                \hline \\[-3mm]

                \end{tabular}

        \end{tabular}

        \end{center}

        $^a$ 80\% completeness limits derived from the recovery rate of artificial sources inserted at empty sky regions.

\end{table*}

\begin{table*}
        \begin{center}
        \caption*{$-$ {\it continued}}
        \label{tab:photometry2}

        {\vskip 1mm}
        \begin{tabular}{c @{\hskip 15mm} c}

                \begin{tabular}{llll}

                \hline \\[-3.3mm]  

                Filter & Telescope & Instrument & Depth$^a$ \\[0mm]

                \hline \\[-3mm]
                XLSS005 \\[-1mm]
                \hline \\[-5mm]  
                \hline \\[-2mm]

                $u$     &   CFHT   &   MegaCam   &   26.0   \\
                $g$     &   CFHT   &   MegaCam   &   26.5   \\
                $r$     &   CFHT   &   MegaCam   &   26.1   \\
                $i$     &   CFHT   &   MegaCam   &   25.8   \\
                $z$     &   CFHT   &   MegaCam   &   25.0   \\
                $R_C$  &   Subaru   &   Suprime-Cam   &   26.0   \\
                $Z_+$  &   Subaru   &   Suprime-Cam   &   24.8   \\
                $J$     &   CFHT   &   WIRCam   &   23.0   \\
                $H$     &   CFHT   &   WIRCam   &   22.5   \\
                $K_s$   &   CFHT   &   WIRCam   &   21.8   \\
                $J$    &   UKIRT   &   WFCAM   &   22.7   \\
                $K$    &   UKIRT   &   WFCAM   &   21.3   \\
                $[3.6]$  &   {\it Spitzer}  &   IRAC   &  24.6    \\
                $[4.5]$  &   {\it Spitzer}  &   IRAC   &  24.4    \\
                $[5.8]$  &   {\it Spitzer}  &   IRAC   &  21.3    \\
                $[8.0]$  &   {\it Spitzer}  &   IRAC   &  21.1    \\[1mm]
                \hline \\[-3mm]

                \hline \\[-3mm]
                RXJ1221 \\[-1mm]
                \hline \\[-5mm]  
                \hline \\[-2mm]

                $B$    &   Subaru   &   Suprime-Cam   &   26.6   \\
                $V$    &   Subaru   &   Suprime-Cam   &   26.1   \\
                $r'$     &   Palomar   &   LFC   &   24.2   \\
                $i'$     &   Palomar   &   LFC   &   24.3   \\
                $z'$     &   Palomar   &   LFC   &   22.8   \\
                $J$    &   UKIRT   &   WFCAM   &   22.4   \\
                $K$    &   UKIRT   &   WFCAM   &   21.9   \\
                $[3.6]$  &   {\it Spitzer}  &   IRAC   &   23.9   \\
                $[4.5]$  &   {\it Spitzer}  &   IRAC   &   23.7   \\[1mm]

                \hline \\[-3mm]

                \end{tabular}

                &

                \begin{tabular}{llll}

                \hline \\[-3mm]  

                Filter & Telescope & Instrument & Depth$^a$ \\[0mm]

                \hline \\[-3mm]
                RXJ1053 \\[-1mm]
                \hline \\[-5mm]  
                \hline \\[-2mm]

                $u$    &   CFHT     &   MegaCam       &   24.8   \\
                $g$    &   CFHT     &   MegaCam       &   25.7   \\
                $r$    &   CFHT     &   MegaCam       &   24.5   \\
                $z$    &   CFHT     &   MegaCam       &   23.6   \\
                $B$      &   Subaru   &   Suprime-Cam   &   26.1   \\
                $V$      &   Subaru   &   Suprime-Cam   &   26.1   \\
                $R_C$    &   Subaru   &   Suprime-Cam   &   25.2   \\
                $R_+$    &   Subaru   &   Suprime-Cam   &   26.4   \\
                $I_+$    &   Subaru   &   Suprime-Cam   &   25.1   \\
                $Z_+$    &   Subaru   &   Suprime-Cam   &   25.5   \\
                $J$   &   UKIRT   &   WFCAM   &   22.3   \\
                $K$   &   UKIRT   &   WFCAM   &   21.7   \\
                $[3.6]$  &   {\it Spitzer}  &   IRAC   &   23.9   \\
                $[4.5]$  &   {\it Spitzer}  &   IRAC   &   23.4   \\
                $[5.8]$  &   {\it Spitzer}  &   IRAC   &   21.7   \\
                $[8.0]$  &   {\it Spitzer}  &   IRAC   &   21.8   \\[1mm]

                \hline \\[-3mm]
                CL1350 \\[-1mm]
                \hline \\[-5mm]  
                \hline \\[-2mm]

                $B$    &   Subaru   &   Suprime-Cam   &   26.5   \\
                $V$    &   Subaru   &   Suprime-Cam   &   25.8   \\
                $R_C$  &   Subaru   &   Suprime-Cam   &   25.1   \\
                $g$    &   CFHT     &   MegaCam       &   24.4   \\
                $r$    &   CFHT     &   MegaCam       &   24.3   \\
                $r'$   &   Palomar  &   LFC           &   25.0   \\
                $i'$   &   Palomar  &   LFC           &   23.5   \\
                $z'$   &   Palomar  &   LFC           &   22.9   \\
                $[3.6]$  &   {\it Spitzer}  &   IRAC   &   23.4   \\
                $[4.5]$  &   {\it Spitzer}  &   IRAC   &   23.4   \\[1mm]

                \hline \\[-3mm]

                \end{tabular}

        \end{tabular}

        \end{center}

        %$^a$ 80\% completeness limits derived from the recovery rate of artificial sources inserted at empty sky regions.

\end{table*}

\nocite{*}
\bibliographystyle{mnras}
\bibliography{bibliography}

\end{document}